\documentclass[preprintnumbers,showpacs,pre,amsmath,amssymb]{revtex4}
\usepackage{mathrsfs}
\usepackage{amssymb}
\usepackage{graphicx}
\usepackage{amsmath}
\usepackage{amssymb}
\usepackage{graphicx}
\usepackage{dcolumn}
\usepackage{bm}
\newcommand{\bra}[1]{\left(#1\right)}
\newcommand{\brb}[1]{\left[#1\right]}

\newcommand{\bre}[1]{\left\{#1\right\}}

\newcommand{\linea}{\noindent\rule{1.0\textwidth}{1pt}}
\newcommand{\emb}{e^{-\beta}}

\begin{document}

\preprint{}

\title{
Stability analysis on the finite-temperature replica-symmetric and
first-step replica-symmetry-broken
cavity solutions of the random vertex cover problem}

\author{Pan Zhang$^{1,3}$}
\author{Ying Zeng$^1$}
\author{Haijun Zhou$^{1,2}$}
\affiliation{$^1$Key Laboratory of Frontiers in Theoretical Physics and
        $^2$Kavli Institute for Theoretical Physics China,
        Institute of Theoretical Physics, Chinese Academy of Sciences,
        Beijing 100190, China}
\affiliation{$^3$Institute of Theoretical Physics, Lanzhou University,
 Lanzhou 730000, China}

\date{\today}

\begin{abstract}
The vertex-cover problem is a prototypical hard combinatorial
optimization problem. It was studied in recent years by physicists
using the cavity method of statistical mechanics. In this paper, the
stability of the finite-temperature replica-symmetric (RS) and the
first-step replica-symmetry-broken (1RSB) cavity solutions of the
vertex cover problem on random regular graphs of finite vertex-degree $K$ are
analyzed by population dynamics simulations. We found that
(1) the lowest temperature for the RS solution to be stable,
$T_{RS}(K)$, is not a monotonic function of $K$, and (2) at relatively
large connectivity $K$ and temperature $T$ slightly below the
dynamic transition temperature $T_d(K)$, the 1RSB
solutions with small but non-negative complexity values are stable.
Similar results are obtained on random Poissonian graphs.
\end{abstract}

 \maketitle

\section{Introduction}
\label{sec:Introduction}

The vertex-cover (VC) problem, which asks to find a set of vertices
of a graph such that the number of vertices in this set is less than
a given value and that each edge of the graph is incident to at
least one of the vertices in the set, is a prototype of NP-complete
problems \cite{GJ79,HW05} with wide range of real-world applications
\cite{BCGRS01,PL01,GEM06}. In the last ten years, the VC problem
defined on the ensemble of large random graphs was extensively
studied using mean-field spin glasses methods, especially the
zero-temperature cavity method \cite{WH01mvfrg, Z03, WZ06, Z05a,
ZMZ07}. It is found that,  when the average connectivity $c$ of the
random graph is such that $c<e= 2.71828\cdots$, the random minimal
vertex-cover (MVC) problem, which corresponds to vertex-covers of
the minimal size for a given random graph, can be described by the
replica-symmetric (RS) cavity theory. In this parameter range,
minimal vertex-covers for a given random graph can be constructed
using a leaf-removal algorithm \cite{BG2001} or by a simple
message-passing warning propagation algorithm \cite{WH01mvfrg}. When
$c > e$, the RS cavity theory is insufficient for the MVC problem,
but the cavity theory at the level of first-step
replica-symmetry-breaking (1RSB) is still able to give accurate
predictions on the average size of minimal vertex-covers
\cite{Z03,WZ06} and the average ground-state entropy \cite{ZZ09}.
Following  the zero-temperature 1RSB cavity theory, a survey
propagation algorithm was used in Ref.~\cite{WZ06} to construct
minimal vertex-covers for single random graphs.

Similar as the mean-field work on the random $K$-satisfiability
problem \cite{MPZ02,MZ02,MP03}, the zero-temperature cavity
calculations of Refs.~\cite{Z03,WZ06} for the VC problem considers
only the energetic effect and ignores completely the entropic
effect. To have a more comprehensive understanding of the random VC
problem, in the present paper a finite temperature $T$ is introduced
into the VC problem. The stabilities of the mean-field RS and 1RSB
cavity solution for the VC problem at finite temperature are
analyzed following earlier works of
Refs.~\cite{KZ07,MRS08,ZK07,ZDEB08,MR03,MPR04,KPW04}. The
$T\rightarrow 0$ limit of these solutions and their stability are
also studied. A similar analysis was recently carried out for the
random $Q$-coloring problem by Krzakala and Zdeborova \cite{KZ07}.
The results reported in this paper suggest that the VC problem and
the $Q$-coloring problem have some important differences. On random
regular graphs with connectivity $K$, for the VC problem we find
that the lowest temperature for the RS solution to be stable,
$T_{RS}(K)$, is not a monotonic function of $K$. The same
non-monotonic behavior is observed for the VC problem on random
Poisson graphs. We also find that,  for random regular graphs with
relatively large connectivity $K$, there is a temperature range
$T_{RS}(K) < T < T_{d}(K)$ in which both a stable RS solution and a
stable 1RSB solution co-exist, where $T_d(K)$ is the dynamical
transition temperature of the system. Although the VC problem is a
spin-glass problem with only two-body interactions, the numerical
results of this work suggest that, for random regular graphs with
relatively large connectivity $K$ and temperature $T \sim T_d(K)$,
the 1RSB cavity solutions with small but non-negative complexity
values may be stable toward further steps of
replica-symmetry-breaking.

The paper is organized as follows. Section~\ref{sec:definitions}
includes some definitions. In Sec.~\ref{sec:ergodic} we consider the RS
solution and its stability.  In Sec.~\ref{sec:1rsb} we consider
the finite temperature 1RSB solution and its type-II stability. We also
compare results obtained by the finite-temperature stability analysis
with those obtained using both the energetic and the entropic
zero-temperature stability
analysis. We conclude this work in Sec.~\ref{sec:conclude}. Some of the
technical details are included in the two appendices of this paper.

\section{Definitions}
\label{sec:definitions}

Two ensembles of random graphs are considered in this paper, namely
Erd{\"{o}}s-Renyi (ER) random graphs \cite{Bollobas01} and regular
random graphs.

An ER random graph ${\cal G}$ has $N$ vertices and
$M= (c/2)N$ different edges, where the edges are chosen completely
random from the set of $N (N-1) /2$ candidate edges between
vertex-pairs. An edge of the graph which links between vertices $i$
and $j$ is denoted by $(i, j)$. The vertex degree $k_i$ of a vertex
$i$ is equal to the number of edges that are linked to vertex $i$.
The mean value of vertex degrees as averaged over all the vertices
of graph ${\cal G}$ is equal to $c$. For a large ER random graph,
the fraction of vertices with a given degree $k$ is given by the
Poisson distribution $f_c(k)\equiv e^{-c} c^k / k!$.
Because of this reason, an ER random graph is also called a Poisson
random graph.

A regular random graph has $N$ vertices and $M = (K/2) N$ edges, with
each vertex having exactly $K$ edges. The $M$ edges in a regular
random graph are randomly connected under the constraint that each
vertex has $K$ edges attached.

A vertex cover of graph ${\cal G}$ is a subset of vertices $U$ that covers
all edges. Here, cover means that for
each  edge $(i,j)$ in the graph at least one of the two end vertices is in
the set $U$.
 We denote the state of each vertex $i$ by a Ising spin
$\sigma_i\in\{ \pm 1 \}$:
$\sigma_i=-1$ if $i\in U$ and $\sigma_i=+1$ if otherwise.
All the vertex covers of  graph
${\cal G}$ form a solution sub-space out of the total number of $2^N$
possible spin configurations.
Each vertex cover
$\{ \sigma_i \} \equiv \{ \sigma_1, \sigma_2, \ldots, \sigma_N \}$
is assigned an energy
\begin{equation}
 \label{eq:E:def}
   E\bra{\{ \sigma_i\}}=\sum_{i=1}^N\delta_{\sigma_i,-1}
\end{equation}
which is equal to the cardinality of the vertex cover.

We introduce a temperature $T$ and weight each vertex cover
by the Boltzmann factor
$e^{-\beta E}$, where $\beta = 1/ T$ is called the inverse temperature.
 The total partition function
$Z$ and the free energy $F(\beta)$ are then defined by
\begin{equation}
 \label{eq:def:z}
  Z \equiv e^{-\beta F(\beta)} =
\sum_{\bre{\sigma_i}} e^{-\beta E\bra{ \bre{\sigma_i} } }
    \prod_{(i,j)} \bigl( 1- \delta_{\sigma_i,1} \delta_{\sigma_j,1} \bigr) \ ,
\end{equation}
and the Gibbs measure for each vertex cover is
\begin{equation}
 \label{eq:pro:def:ft}
\mathcal{P}\bra{\{ \sigma_i \}}=\frac{1}{Z}e^{-\beta E\bra{ \{
\sigma_i \} }} \prod_{(i,j)} \bigl( 1-\delta_{\sigma_i,1}
\delta_{\sigma_j,1} \bigr) \ .
\end{equation}
In Eqs.~(\ref{eq:def:z}) and (\ref{eq:pro:def:ft}), the term
$\prod_{(i,j)} ( 1 - \delta_{\sigma_i,1} \delta_{\sigma_j,1})$ is
equal to unity or zero depending on whether the spin configuration
corresponds to a vertex-cover or not. Only vertex-covers contribute
to the free-energy of the system.
The $T\rightarrow 0$ limit of Eq.~(\ref{eq:def:z}) corresponds
 to the MVC problem.
In this case, only those ground-state solutions have non-zero Gibbs measure.

Under the Gibbs measure Eq.~(\ref{eq:pro:def:ft}) the marginal probability
 $\pi_i$ of a vertex $i$
being covered is expressed as
\begin{equation}
\label{eq:def:pp}
 \pi_i=\sum_{\{\sigma_j\}} \mathcal{P}\bra{\{ \sigma_j \}}
 \delta_{\sigma_i,-1} \ .
\end{equation}
$\pi_i$ is called the cover ratio of vertex $i$. A direct computation
of the cover ratios $\pi_i$ is
difficult for large random graphs, but approximate values for $\pi_i$
can be obtained using the
cavity method.

\section{Stability of the Replica Symmetric Cavity Theory}
\label{sec:ergodic}

\subsection{The replica-symmetric cavity equations at finite temperatures}

According to the replica-symmetric cavity theory \cite{MP01}, the free energy
at inverse temperature $\beta$ can be calculated by
\begin{equation}
    \label{eq:rs:F}
        F(\beta) =\sum_{i \in {\cal G}} \Delta F_i
        -\sum_{(i,j) \in {\cal G}} \Delta F_{(i,j)} \ ,
\end{equation}
where $\Delta F_{i}$ and $\Delta F_{(i,j)}$ are, respectively,
 the free energy shift
 due to the addition of vertex $i$ and edge $(i,j)$.
The free energy expression Eq.~(\ref{eq:rs:F}) corresponds to the zeroth-order
 term of a loop series expansion for the total partition function
Eq.~(\ref{eq:def:z}) \cite{Chertkov-Chernyak-2006b}.
The set of nearest-neighbors for a vertex $i$ is denoted as $\partial i$.
 Because of the locally tree-like structure of
a random graph ${\cal G}$, in the absence of vertex $i$, the length
of the shortest paths between two vertices $j, k$ in the set
$\partial i$ diverges logarithmically with the graph size $N$. It is
then
 assumed that in the absence of
vertex $i$ the spin values on the vertices of the set $\partial i$
are mutually independent. Under this Bethe-Peierls approximation, the free
energy shift associated with the addition of vertex $i$ is expressed as
\begin{equation}
    \label{eq:rs:deltaFi}
    \Delta F_{i}=-\frac{1}{\beta}\log\Bigl( e^{-\beta}+\prod_{j
        \in\partial i} \pi_{j|i}(\beta) \Bigr) \ ,
\end{equation}
where $\pi_{j|i}(\beta)$ is the probability of vertex $j$ being covered in
the absence of vertex $i$.  In Eq.~(\ref{eq:rs:deltaFi}), the term
$e^{-\beta}$ corresponds to vertex $i$ being covered ($\sigma_i=-1$),
while the term $\prod_{j\in \partial i} \pi_{j|i}$ corresponds to vertex $i$
being uncovered (then all the neighbors of $i$ need to be covered).
Under the same Bethe-Peierls approximation, the free energy shift
$\Delta F_{(i,j)}$ is expressed as
\begin{equation}
    \label{eq:rs:deltaFij}
    \Delta F_{(i,j)}=-\frac{1}{\beta}\log\Bigl(1- \bigl(1- \pi_{i|j}(\beta)
     \bigr) (1-\pi_{j|i}(\beta)\bigr) \Bigr) \ .
\end{equation}

The free energy $F$ as expressed by Eq.~(\ref{eq:rs:F}) is a functional of
the $2 M$ cavity probabilities  $\{ \pi_{j|i} \}$, two on each edge $(i,j)$.
At equilibrium, the free energy $F$ should reach a minimal value.
Then the variational condition
\begin{equation}
    \label{eq:rs:F:var}
    \frac{\delta F}{\delta \pi_{j|i}} = 0
\end{equation}
leads to the following iterative equation for each cavity probability
$\pi_{j|i}$:
\begin{equation}
    \label{eq:rs:iter}
    \pi_{j|i} = \mathcal{F}_{RS}\bra{\bre{\pi_{k|j}}}=\frac{e^{-\beta}}
    {e^{-\beta}+\prod_{k \in \partial j \backslash i} \pi_{k|j}(\beta) } \ ,
\end{equation}
where $\partial j \backslash i$ denotes the remaining set after vertex $i$ is
removed from set $\partial j$. When a fixed point
is reached for the set of iterative equations Eq.~(\ref{eq:rs:iter}), the mean
energy $\langle E \rangle$ and entropy $S$ of the system are then
calculated according to
\begin{eqnarray}
     \langle E \rangle  & = &\frac{ {\rm d} \beta F }{{\rm d} \beta}
           = \sum_i \frac{e^{-\beta}}{e^{-\beta}+\prod\limits_{j \in
        \partial i} \pi_{j|i}(\beta) } \ ,
    \label{eq:rs:E}  \\
     S & = & \beta \bigl( \langle E \rangle -F \bigr) \ . \label{eq:rs:S}
\end{eqnarray}

For a single graph ${\cal G}$, we denote by $\mathcal{P}_{RS}( \pi
)$ the probability of observing a cavity probability with value
$\pi_{j|i}=\pi$, namely
\begin{equation}
    \label{eq:rs:Pcv}
    \mathcal{P}_{RS}(\pi)= \frac{1}{2 M} \sum\limits_{(i,j)\in {\cal G}}
    \Bigl(\delta\bigl(\pi_{j|i}(\beta)-\pi\bigr)
    +\delta\bigl(\pi_{i|j}(\beta)-\pi\bigr) \Bigr)
    \ ,
\end{equation}
where $\delta(x)$ is the Dirac delta function. When the size $N$ of a
random graph ${\cal G}$ is sufficiently large,
the probability distribution
$P(\pi)$ becomes independent of the detailed connection pattern of the
graph. It only depends on the vertex degree profile of the graph and the
inverse temperature $\beta$. We can write down the following self-consistent
equation for $P(\pi)$:
\begin{equation}
    \label{eq:rs:pp}
    \mathcal{P}_{RS}(\pi)= p_{nn}(1)
    \delta\Bigl(\pi-\frac{e^{-\beta}}{e^{-\beta}+1} \Bigr) +
    \sum\limits_{k=1}^\infty p_{nn}(k+1)
    \int \prod_{j=1}^{k} \Bigl[ {\rm d} \pi_j \mathcal{P}_{RS}(\pi_j) \Bigr]
    \delta\Biggl( \pi - \frac{e^{-\beta} }{e^{-\beta} + \prod_{j=1}^k
    \pi_j} \Biggr) \ .
\end{equation}
In Eq.~(\ref{eq:rs:pp}), $p_{nn}(k+1)$ is the probability that a randomly
chosen nearest-neighbor of a vertex have vertex degree $k+1$.

\begin{figure}
\includegraphics[width=0.47\textwidth]{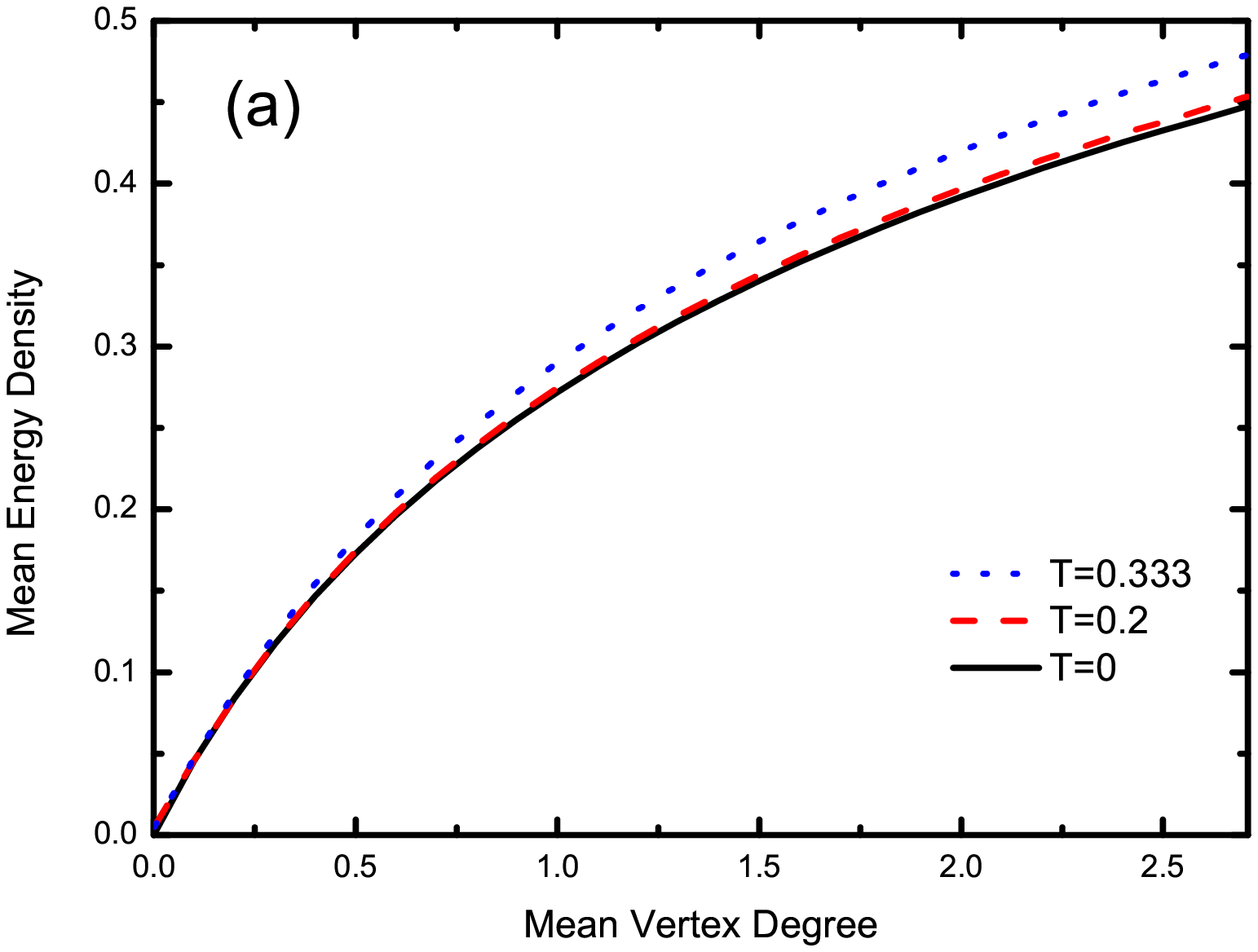}\hspace{0.5cm}
\includegraphics[width=0.47\textwidth]{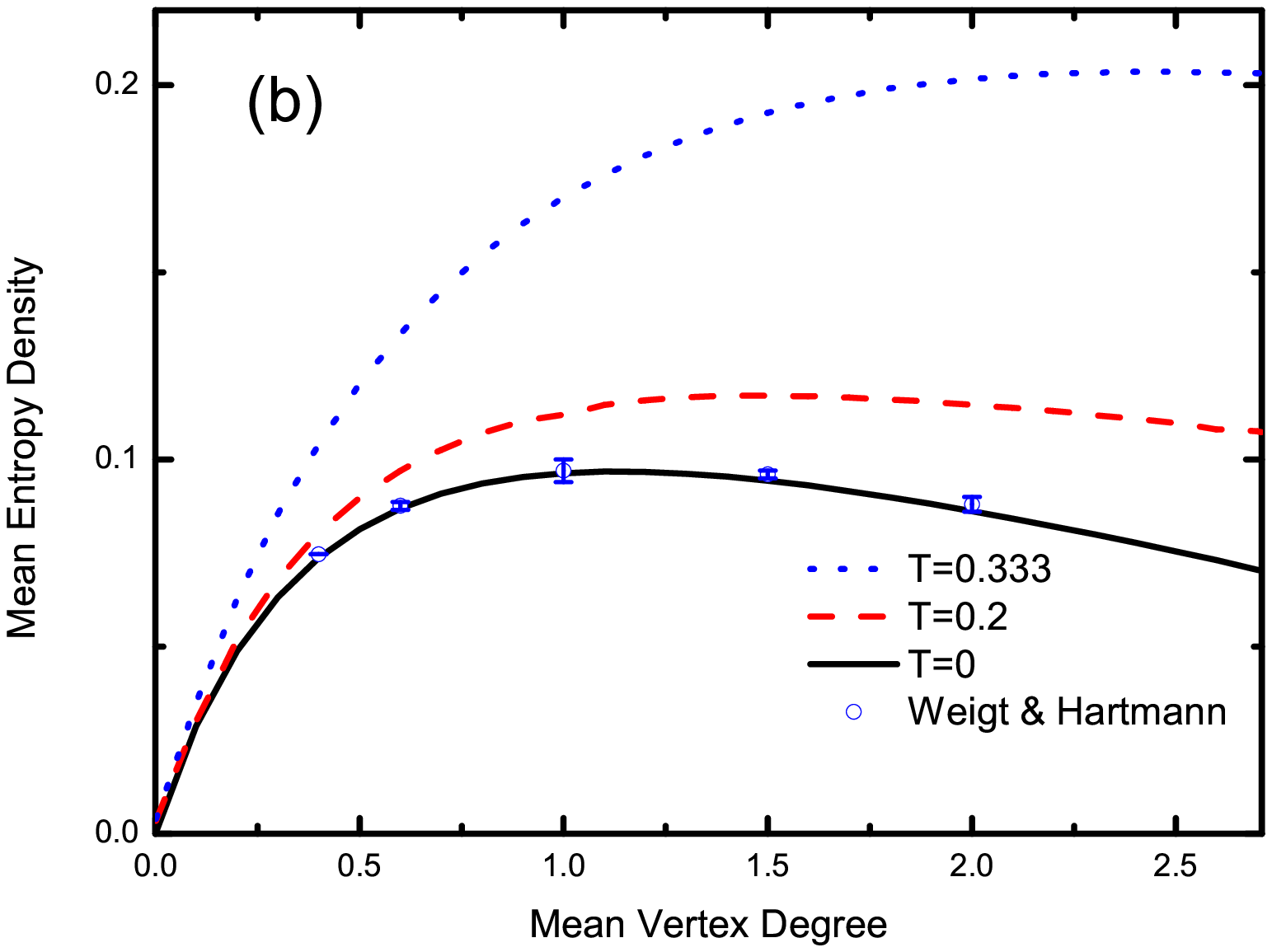}
\caption{ \label{fig:ft:zt:es}
(Color Online) Mean vertex-cover
energy density and entropy density as a function of the mean vertex
degree of the ER random graph. Different curves correspond to
different temperature $T$. The circular symbols are
simulation results of Ref.~\cite{WH01mvfrg}. }
\end{figure}

For ER random graphs, $p_{nn}(k+1)= (k+1) f_c(k) / c = f_c(k)$,
i.e., it is also a Poisson distribution. A fixed-point solution for
Eq.~(\ref{eq:rs:pp}) can be obtained by population dynamics
simulation \cite{MP01}. In terms of the cavity probability
distribution $\mathcal{P}_{RS}(\pi)$, the densities for the free
energy, mean energy, and entropy can be re-written as
\begin{eqnarray}
    f \equiv F/N &=& -\frac{1}{\beta} f_c(0) \log\bigl(e^{-\beta}+1\bigr)
    -\frac{1}{\beta}\sum_{k=1}^\infty f_c(k) \int
    \prod_{j=1}^k \Bigl[{\rm d} \pi_j \mathcal{P}_{RS}(\pi_j) \Bigr]
    \log\Bigl(e^{-\beta}+\prod_{j=1}^k \pi_j \Bigr)
    \nonumber\\
     & & +\frac{c }{2\beta} \int {\rm d} \pi_i \mathcal{P}_{RS}(\pi_i)
    {\rm d} \pi_j \mathcal{P}_{RS}(\pi_j) \log\bigl(1-(1-\pi_i)(1-\pi_j)\bigr) \ ,
    \label{eq:rs:f} \\
    \bar{e}\equiv \langle E \rangle /N
    &=&f_c(0) \frac{e^{-\beta}}{e^{-\beta}+1}
    + \sum\limits_{k=1}^\infty f_c(k)
    \int \prod\limits_{j=1}^k \Bigr[
    {\rm d} \pi_j \mathcal{P}_{RS}(\pi_j) \Bigr]
    \frac{e^{-\beta}} {e^{-\beta}+\prod_{j=1}^{k}
    \pi_j}  \ ,
    \label{eq:rs:e} \\
    s\equiv S/N &=&\beta (\bar{e}-f) \ .
    \label{eq:rs:s}
\end{eqnarray}

The mean energy and entropy density of the vertex-cover
problem on ER random graphs of mean degree $c < 2.7183$ are shown in Fig.~\ref{fig:ft:zt:es}.
At a given value of $\beta$, the mean
energy density increases continuously with the mean vertex degree $c$.
On the other hand, the mean entropy first increases with $c$ when $c$ is small
and then decreases with $c$ when $c$ exceeds certain temperature-dependent value.

For regular random graphs, $p_{nn}(k+1)$ in Eq.~(\ref{eq:rs:pp}) is
expressed as $p_{nn}(k+1) = \delta_{k}^{K-1}$. In the
replica-symmetric cavity theory, it is therefore assumed that the
cavity cover ratio distribution $\mathcal{P}_{RS}(\pi)$ is a Dirac
delta function $\mathcal{P}_{RS}(\pi)= \delta ( \pi - \pi_*)$, with
$\pi_*$ determined by
\begin{equation}
 \label{eq:rs:rer:iter}
\pi_* e^{-\beta}+\pi_*^{K}-e^{-\beta}=0 \ .
\end{equation}
The mean free energy density and energy density are calculated by
\begin{eqnarray}
 f &=&-\frac{1}{\beta}\log(e^{-\beta}+\pi_*^{K})+\frac{K}{2\beta}\log(2\pi_*-\pi_*^2) \ ,  \\
 \bar{e} &=& \frac{e^{-\beta}}{e^{-\beta}+\pi_*^{K}}  \ .
\end{eqnarray}

\subsection{The entropic zero-temperature limit}
\label{sec:rs:zerotemperature}

The energetic zero-temperature limit of the RS cavity theory is very easy to
implement. In this limit, one only interests in whether a given
vertex is always uncovered among all the MVCs, and
a warning propagation algorithm can be constructed for the
vertex-cover problem in this limit \cite{WZ06}.
In this subsection, we study the entropic zero-temperature limit so that the
entropy of MVCs can also be calculated.

It is helpful to define two auxiliary parameters $\eta_i(\beta)$ and
$\eta_{j|i}(\beta)$ through
\begin{eqnarray}
    \eta_i(\beta) &=&
    \frac{1}{\beta} \log\Bigl(\frac{\pi_i(\beta)}{1-\pi_i(\beta)}
        \Bigr)  \ ,
    \label{eq:def:eta} \\
    \eta_{j|i}(\beta) &=&
    \frac{1}{\beta} \log\Bigl(\frac{\pi_{j|i}(\beta)}{1-\pi_{j|i}(\beta)}
    \Bigr) \ .
\end{eqnarray}
The physical meanings of $\eta_i$ and $\eta_{j|i}$ are obvious:
$\beta \eta_i= \log[\pi_i/(1-\pi_i]$ is the log-likelihood of vertex $i$ being
covered, and $\beta \eta_{j|i}$ is the log-likelihood of vertex $j$ being
covered in the absence of vertex $i$. The iterative equation (\ref{eq:rs:iter})
can be rewritten as
\begin{equation}
    \label{eq:rs:iter-eta}
    \eta_{j|i}(\beta) = -1 + \sum\limits_{k\in \partial j \backslash i}
    \frac{1}{\beta}\log\bigl(1+ e^{-\beta \eta_{k|j}}\bigr) \ .
\end{equation}

At $T\rightarrow 0$ ($\beta \rightarrow +\infty$)
we assume that
\begin{equation}
    \label{eq:rs:eta-infty}
    \eta_{j|i} = m_{j|i} + \frac{r_{j|i}}{\beta} \ ,
\end{equation}
with $m_{j|i}$ being an integer and $r_{j|i}$ being a finite real value. Then
From Eq.~(\ref{eq:rs:iter-eta}) we get the iteration equations for
$m_{j|i}$ and $r_{j|i}$:
\begin{eqnarray}
    m_{j|i} &=& -1 + \sum\limits_{k\in \partial j \backslash i}
    \Theta(-m_{k|j}) \ ,
    \label{eq:rs:eta-infty:m} \\
    r_{j|i} &=& \sum\limits_{k\in \partial j \backslash i}
    \Bigl[\bigl(1-\Theta(|m_{k|j}|)\bigr) \log(1+e^{-r_{k|j}})
    -\Theta(-m_{k|j}) r_{k|j} \Bigr] \ ,
    \label{eq:rs:eta-infty:r}
\end{eqnarray}
where $\Theta(x)$ is the Heaviside function defined by $\Theta(x)=1$ for $x>0$
and $\Theta(x)=0$ for $x\leq 0$.

At the limit of $\beta \rightarrow \infty$, the free energy
Eq.~(\ref{eq:rs:f}), energy Eq.~(\ref{eq:rs:e}), and entropy
Eq.~(\ref{eq:rs:s}) can all be expressed in terms of $\{ m_{j|i},
r_{j|i}\}$. From these expressions, the mean ground-state energy and
entropy densities of the VC problem on an ER random graph can easily
be evaluated by population dynamics. The theoretical predictions on
the mean energy and entropy density of the MVC problem are also
shown in Fig.~\ref{fig:ft:zt:es}, together with the simulation
results of Weigt and Hartmann \cite{WH01mvfrg}. For mean
connectivity $c < 2.7183$ the agreement between theory and
simulation results is good. In the population dynamics simulation, we
have noticed that when $c>2.7183$, the amplitude of some $r_{j|i}$ values
approaches infinity when $m_{j|i}=0$.  Such type of divergence then lead
to a negative
value for the entropy density of the MVC problem (see also
Ref.~\cite{ZZ09}). As we will discuss in the
next subsection, for ER random graphs with $c > 2.7183$, the zero-temperature
RS cavity theory is no longer valid and a more advanced mean-field theory is
needed.

We notice that, the divergence of the residue fields $r_{j|i}$ as
observed for the random MVC problem does not occur in the random maximal
matching problem \cite{ZM06}. For the random maximal matching problem,
the RS cavity theory is stable at any temperature.

\subsection{Stability of the replica-symmetric solution}
\label{sec:rs-stability}

At low enough temperatures and/or high mean vertex degrees, the
Bethe-Peierls approximation used in the RS cavity equations is no
longer valid. Then the RS cavity theory becomes unstable to higher
levels of replica-symmetry-breaking. The stability of the RS cavity
equations can be checked by studying the point-to-set correlations
in the graph \cite{KMRSZ07,MM06,MS06}. If these correlations do not
decay to zero at large distances, then non-trivial solutions exist
for the one-step replica-symmetry-breaking (1RSB) cavity equations
at Parisi parameter $m=1$. The dynamical transition temperature
$T_d$, which is defined by the critical temperature where
point-to-set correlation begin to diverge, can be checked using 1RSB
equations at $m=1$(see Appendix~\ref{sec:appendixb} for a detailed
calculation of $T_d$).

A easier way to check the validity of the RS assumption is to study the
local stability of the RS solution. This local stability analysis
leads to a threshold temperature $T_{RS}$.
However, the local stability of the RS solution is a
necessary but not a sufficient condition for RS correctness and in
general, $T_{RS}\leq T_{d}$. In this paper, the way of checking the
local stability of the RS solution is to study the spin-glass
susceptibility \cite{ZDEB08,ZM06} as defined by
\begin{equation}
    \label{eq:rs:sta:def:kafa}
    \chi_{SG}=\frac{1}{N}\sum_{i\neq j} \langle \sigma_i\sigma_j
    \rangle_c^2 \ ,
\end{equation}
where $\langle \sigma_i \sigma_j \rangle_c \equiv
\langle \sigma_i \sigma_j \rangle - \langle \sigma_i \rangle
\langle \sigma_j \rangle$ is the connected correlation between
vertex $i$ and vertex $j$. The above equation can be re-expressed as
\begin{equation}
    \label{eq:rs:sta:kafa}
    \chi_{SG} = \frac{2}{N} \Bigl[
    N_1 \overline{ \langle \sigma_i \sigma_{j(1)} \rangle_c^2}
    + N_2 \overline{ \langle\sigma_i \sigma_{j(2)} \rangle_c^2}
    +\ldots
    +N_d \overline{\langle \sigma_i \sigma_{j(d)} \rangle_c^2}
    +\ldots \Bigr] \ ,
\end{equation}
where $N_d$ is the total number of vertex-pairs of distance
(minimum path length) $d$ in the graph ${\cal G}$;
$\sigma_{j(d)}$ denotes a vertex $j$ which is separated from vertex $i$ by
a distance $d$; and $\overline{\langle \sigma_i \sigma_{j(d)}
\rangle_c^2}$ denotes the mean value of $\langle \sigma_i \sigma_{j(d)}
\rangle_c^2$ as averaged over all the $N_d$ vertex-pairs $(i, j)$ of
distance $d$. For a large ER random graph with mean
connectivity $c$, $N_d = N (c^d/2)$ when $d$ is much smaller
than the length
of a typical loop in the graph ($d < \log_c N$), while for a large regular
random graph with vertex degree $K$, the scaling is $N_d = N K (K-1)^{d-1}/2$.
On the other hand, using the locally tree-like property of a random graph
$\mathcal{G}$, it can be shown that
\begin{equation}
    \label{eq:correlation_d}
    \overline{ \langle \sigma_i \sigma_{j(d)} \rangle_c^2}
    \propto  \lambda^{(d-1)} \ ,
\end{equation}
where
\begin{equation}
    \label{eq:lambda}
    \lambda(\beta)= \overline{\Biggl[
    \frac{ \partial \pi_{j|i} }{\partial \pi_{k|j}} \Biggr]^2
    }
    =
    \overline{
    \Biggl[
    \frac{e^{-\beta} \prod_{l\in \partial j \backslash i, k}
    \pi_{l|j}(\beta)}{
\bigl(e^{-\beta}+\prod_{l\in \partial j\backslash i} \pi_{l|j}(\beta)
\bigr)^2}
    \Biggr]^2 }
 \ .
\end{equation}
In the above equation, the overline means averaging over all the paths
$k\rightarrow j\rightarrow i$ of
length two in the graph $\mathcal{G}$.

\begin{figure}
\includegraphics[width=0.47\textwidth]{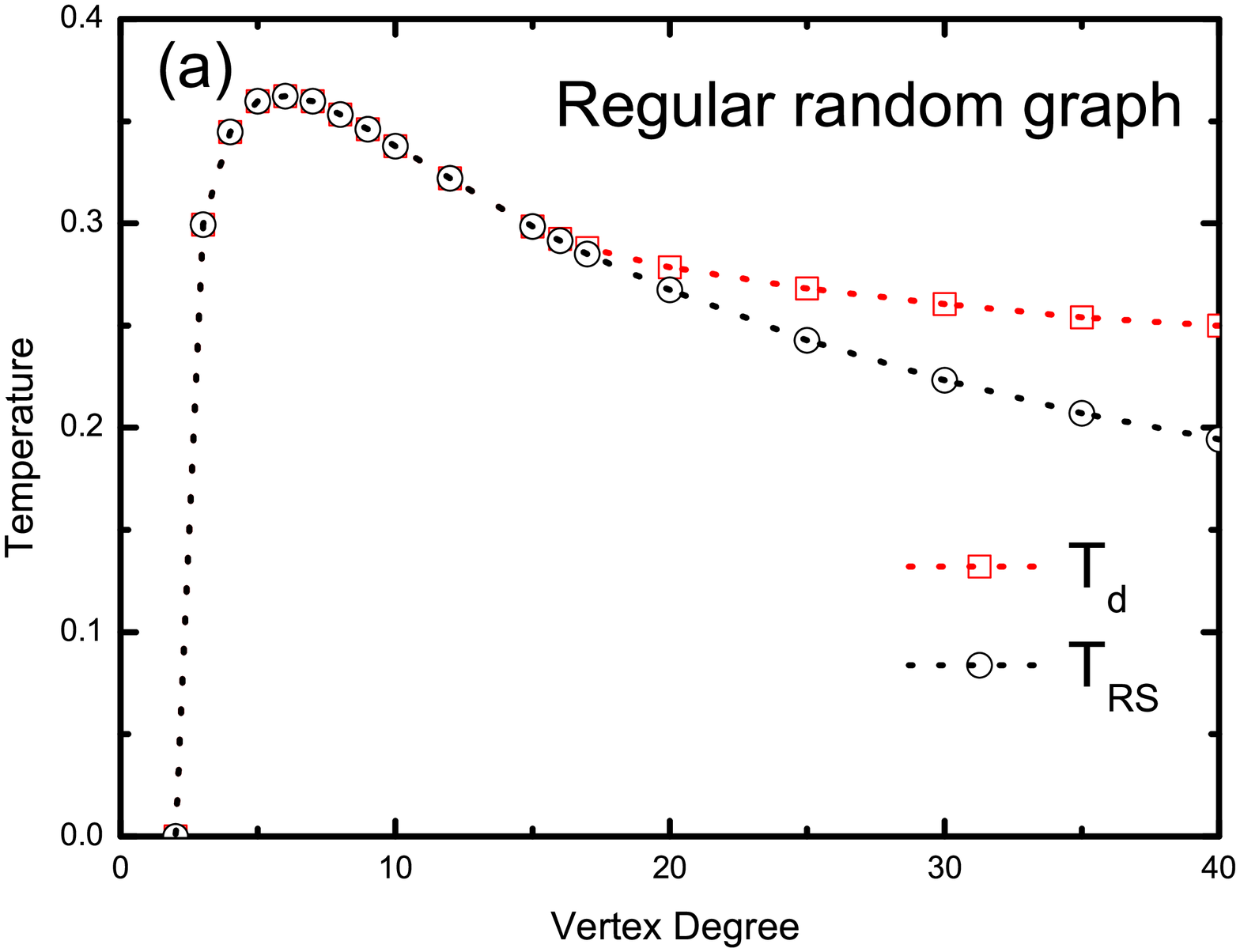}\hspace{0.5cm}
\includegraphics[width=0.47\textwidth]{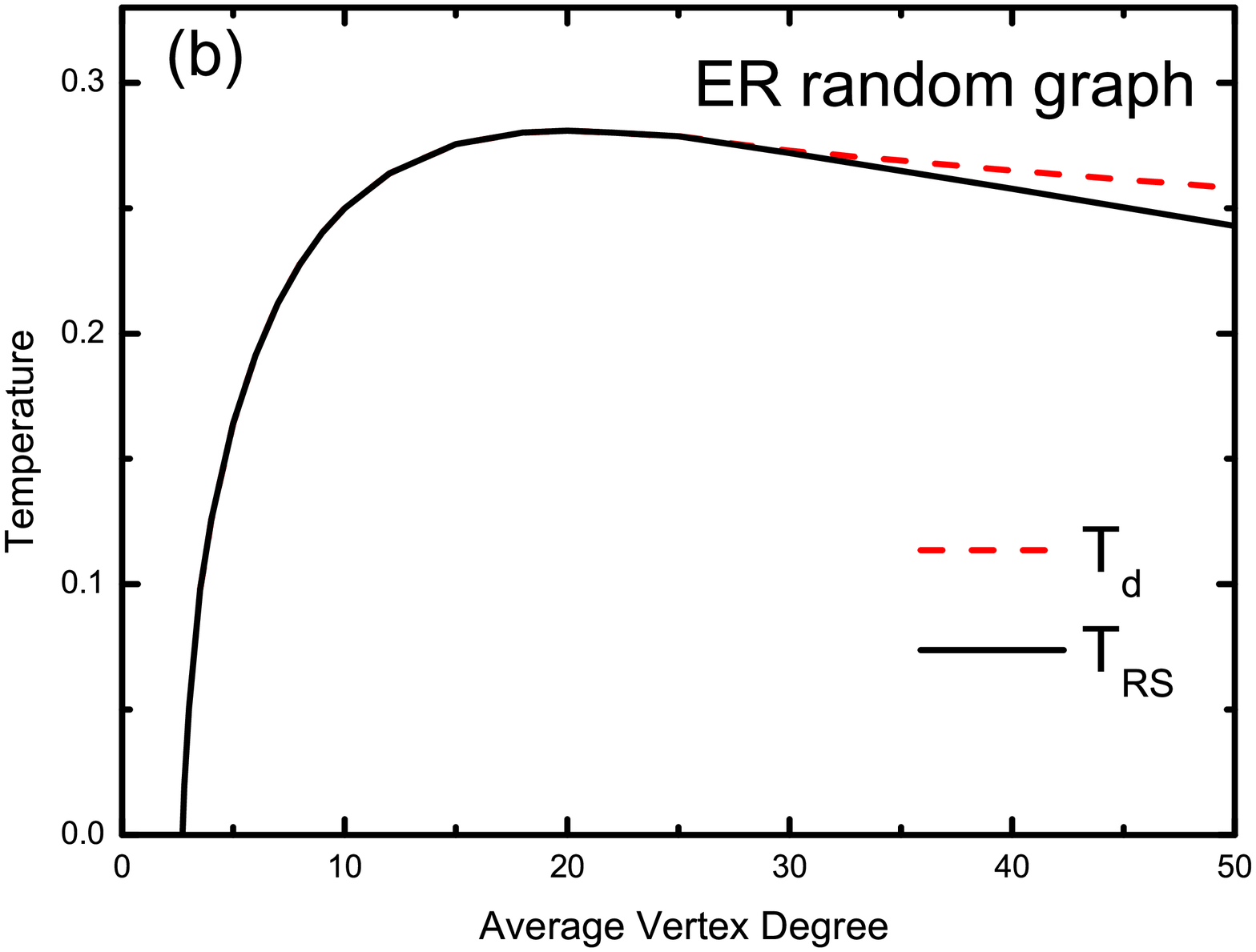}
\caption{
 \label{fig:graph:rstab}
(a) RS local instability temperature $T_{RS}$ and dynamical transition
temperature $T_d$ of the vertex-cover problem on regular random graphs (a)
and ER random graphs (b). When $T<T_{RS}$, RS solution become locally unstable;
When $T<T_d$, 1RSB solution at $m=1$ has non-trivial solution.}
\end{figure}

For a regular random graph, the spin-glass susceptibility $\chi_{SG}$ remains
finite in the thermodynamic limit of $N\rightarrow \infty$ if and only if
$(K-1) \lambda(\beta) < 1$. This condition is re-expressed as
\begin{equation}
\label{eq:rs:stab:rer}
(K-1)\Biggl[ \frac{e^{-\beta}\pi_*^{K-2}}{\bra{e^{-\beta}+\pi_*^{K-1}}^2}
\Biggr]^2<1 \ , \end{equation}
where $\pi_*$ is the solution of Eq.(\ref{eq:rs:rer:iter}).

The local stability boundary for the RS cavity theory as predicted
by Eq.~(\ref{eq:rs:stab:rer}) and the dynamical transition
temperature $T_d$ are shown in Fig.~\ref{fig:graph:rstab}a. At
$K=2$, the RS solution is locally stable at any temperature. When $K
\geq 3$ the RS solution is only stable at temperatures $T > T_{RS}$.
The critical temperature $T_{RS}$ is not a monotonic function of
graph degree $K$ but rather has a maximal value at $K=6$.  Such a
re-entrant behavior is also observed for random ER graphs
(Fig.~\ref{fig:graph:rstab}b). It is not yet clear why the RS
solution of the VC problem on a random regular graph of $K=6$ is the
most easiest to be unstable. At small values of $K$, $T_d = T_{RS}$
. But when $K\geq 16$, $T_d>T_{RS}$. The 1RSB solution at $m=1$ (see
the next section and Appendix~\ref{sec:appendixb}) begins to have a
non-trivial solution at $T = T_d$, suggesting that the configuration
space of the system starts to splitting into many Gibbs pure states.
In the temperature region $T_{RS}<T<T_d$, although the RS solution
still remains locally stable, it does not correctly describe the
property of the system. In our numerical solutions, we find that the
Kauzmann temperature $T_K$, which corresponds to zero complexity
($\Sigma\bra{m=1}=0$, see next section), is always equal to $T_d$.

For an ER random graphs, the convergence condition for the
spin-glass susceptibility $\chi_{SG}$ is $c \lambda(\beta) < 1$. For
this ensemble of graphs, we can not determine the stability boundary
analytically. Instead, we iterate a stability parameter
$\Delta_{i|f}$ in population using population dynamics, where
\begin{equation}\label{eq:stab:rs:iter}
 \Delta_{i|f}=\sum_{j\in\partial i\backslash f}
    \left[ \frac{e^{-\beta\prod_{j'\neq j}\pi_{j'|i}}}{\bra{e^{-\beta}
    +\prod_{j\in\partial i\backslash f}\pi_{j|i}^2}}\right]^2\Delta_{j|i}.
\end{equation}

If $\Delta_{j|f}<1$ after iterating for a long enough time, the RS
solution is then locally stable; otherwise it is locally unstable.
The local stability boundary for the RS cavity solution is shown in
Fig.~\ref{fig:graph:rstab}b. When $c<2.7183$ the RS solution is
always stable and 1RSB equation at $m=1$ has trivial solution. When
$c>e$, the RS solution is stable at high temperatures and becomes
unstable blow a threshold temperature $T_{RS}(c)$. The threshold
temperature $T_{RS}(c)$ has a maximal value at mean vertex-degree $c
\approx 20$. Similar with those of regular random graphs,
$T_d=T_{RS}$ at relative small average connectivity and $T_d$
becomes larger than $T_{RS}$ when $c>=30$.

\subsection{Infinite-connectivity limit}

When the connectivity ($c$ for ER graphs and $K$ for regular random graphs)
is large, we see from Eq.~(\ref{eq:rs:iter}) that $\pi_{j| i}$
should be very close to $1$. In the case of regular random graphs,
the solution at the $K\to \infty$ limit of Eq.~(\ref{eq:rs:rer:iter})
has the following property
\begin{equation}
\lim_{K\to\infty}\pi_\star^K=0 \ , \; \; \;
\lim_{K\to\infty}\pi_\star = 1 \ .
\end{equation}
At this limit, the free energy density and mean energy density both equal to
unity, and the entropy density is zero.
This means that  when the connectivity goes to infinity, there is only one
vertex cover for the graph which includes all the vertices.
At $K\rightarrow \infty$, the local stability condition for the RS solution
is
\begin{equation}
    \lim_{K\to\infty}\pi_\star^{2K-4}\bra{K-1}e^{2\beta}<1 \ ,
\end{equation}
which is satisfied at any finite temperature. Therefore the RS solution is
locally stable at any finite temperature for an infinitely connected regular
random graph.

For ER random graphs, we do not have an analytical expression for the
large $c$ limit, but we have checked by
population dynamics simulations that
the results are the same as those in regular random graphs: the cover
ratio and the energy density both go to unity,
the entropy density goes to zero and the RS
solution is locally stable at any finite temperature.

\section{Stability of the first-step replica-symmetry-broken cavity solution}
\label{sec:1rsb}

\subsection{1RSB solution at finite temperatures}

When the RS mean-field solution to the random vertex-cover problem is unstable,
one can try to describe the system using the first-step
replica-symmetry-breaking (1RSB) spin-glass theory.
In the 1RSB theory, the configuration space of the system is divided into
many sub-spaces or macrostates. Each macrostate $\alpha$ has a free energy
$F_\alpha(\beta)$ and its contribution to the statistical property of the
system is weighted by a Boltzmann factor $\exp\bigl(-y F_\alpha(\beta)\bigr)$,
where $y$ is the adjustable inverse temperature at the
level of macrostates. The ratio $m \equiv  y/ \beta = y T $ is called the
Parisi parameter. The grand free energy density (also called the replicated
free energy density)  $g(y,\beta)$ is defined as
\begin{equation}
 \label{eq:1rsb:g:def}
g(y, \beta) = - \frac{1}{ N y}
\log\Bigl( \sum\limits_{\alpha} \exp\bigl(- y F_\alpha(\beta) \bigr) \Bigr)
 = - \frac{1}{N y} \log\Bigl( \int {\rm d}
 f e^{-N y f + N \Sigma(f)} \Bigr) \ ,
\end{equation}
where $f$ denotes the free energy density of a macrostate and
$\exp(N \Sigma(f) )$ is the density of macrostates
with free energy density $f$. The quantity $\Sigma(f)$ is called the
complexity. Taking the $N\to\infty$ limit, at saddle-point we have
    \begin{equation}
         \label{eq:1rsb:sp}
        g(y,\beta)=\min\limits_{f} \bigl[f - \Sigma(f)/y \bigr] \ .
    \end{equation}
The macrostates with the lowest free energy density $f_0(\beta)$ corresponds
to the point of zero complexity, $\Sigma\bigl( f_0(\beta) \bigr)=0$.
Depending on the value of the parameter $y$ (or equivalent
the Parisi parameter $m$) the free energy density $f$ of the macrostates
which contribute to the grand free energy density $g(y, \beta)$
is determined by
    \begin{equation}
        \label{eq:lrsb:cfy}
        \frac{ {\rm d} \Sigma(f) } { {\rm d} f} = y  \ .
    \end{equation}

In the 1RSB cavity theory, the order parameter is no longer the cover ratio
$\pi_{j|i}$ but the distribution profile $Q_{j|i}(\pi_{j|i})$ of $\pi_{j|i}$
over all the macrostates. Equation~(\ref{eq:rs:iter}) is generalized into
    \begin{equation}
         \label{eq:1rsb:iter}
        Q_{j|i}(\pi_{j|i} ) = \mathcal{F}_{1RSB}(\{Q_{k|j}\}) =
        \frac{1}{Z_{j|i} }\Bigl[ \int\prod_{k\in\partial j \backslash i}
        {\rm d} \pi_{k|j} Q_{k|j}(\pi_{k|j}) \Bigr]
        e^{-y\Delta F_{j|i}} \delta \Bigl( \pi_{j|i} -
     \frac{e^{-\beta}}{e^{-\beta} +
        \prod_{k\in \partial j\backslash i} \pi_{k|j} } \Bigr) \ ,
    \end{equation}
where $Z_{j|i}$ is a normalization factor, and the expression for
$\Delta F_{j|i}$ is
    \begin{equation}
        \Delta F_{j|i} = -\frac{1}{\beta} \log\bigl( e^{-\beta} +
       \prod\limits_{k\in \partial j \backslash i} \pi_{k|j}(\beta) \bigr) \ .
    \end{equation}
At given inverse temperatures $\beta$ and $y$,  a fixed-point solution
$\{ Q_{j|i}(\pi_{j|i}) \}$ of  Eq.~(\ref{eq:1rsb:iter}) for a given random
graph can be obtained by population dynamics.
The corresponding grand free energy density $g$, mean free-energy density
$\langle f \rangle$ (averaged over all the macrostates), and complexity can be
obtained by the following equations:
    \begin{eqnarray}
         g  &= & \frac{1}{N} \sum\limits_{i} \Delta G_i -
            \frac{1}{N} \sum\limits_{(i,j)} \Delta G_{i j} \ ,
     \label{eq:1rsb:g} \\
        \langle f \rangle &=& \frac{1}{N}
     \sum\limits_{i} \langle \Delta F_i \rangle
        - \frac{1}{N} \sum\limits_{(i,j)} \langle \Delta F_{i j} \rangle  \ ,
     \label{eq:1rsb:F} \\
        \Sigma & = & y( \langle f \rangle - g) \ .
    \end{eqnarray}
In the above equations, $\Delta G_i$ and $\Delta G_{i j}$ are, respectively,
the shift of the grand free energy of the system due to the addition of a
vertex $i$ and an edge $(i,j)$:
    \begin{eqnarray}
    \Delta G_i &=& -\frac{1}{y}\log\Bigl[\int\prod_{j\in\partial i}
     {\rm d} \pi_{j|i}
            Q_{j|i}(\pi_{j|i}) e^{-y \Delta F_{i}} \Bigr] \ , \\
        \Delta G_{i j} &=&-\frac{1}{y}\log\Bigl[\int {\rm d} \pi_{j|i}
     {\rm d} \pi_{i|j}
            Q_{j|i}(\pi_{j|i}) Q_{i|j}(\pi_{i|j}) e^{-y \Delta F_{i j}}
     \Bigr] \ ;
    \end{eqnarray}
and $\langle \Delta F_{i} \rangle$ and $\langle \Delta F_{i j} \rangle$ are,
respectively, the mean value of the changes $\Delta F_{i}$ and
$\Delta F_{i j}$ over all the macrostates:
    \begin{eqnarray}
        \langle \Delta F_i \rangle &= &
        \frac{\int \prod_{j\in\partial i}
            {\rm d} \pi_{j|i} Q_{j|i}(\pi_{j|i}) \Delta F_i e^{-y \Delta F_i} }
             {\int\prod_{j\in\partial i}{\rm d} \pi_{j|i} Q_{j|i}(\pi_{j|i})
            e^{-y \Delta F_i} } \ ,  \\
        \langle \Delta F_{i j} \rangle &=&
        \frac{\int {\rm d} \pi_{j|i} {\rm d} \pi_{i|j} Q_{j|i}(\pi_{j|i})
     Q_{i|j}(\pi_{i|j})
            \Delta F_{i j}e^{-y\Delta F_{i j}}}
        {\int {\rm d}\pi_{j|i} {\rm d}\pi_{i|j} Q_{j|i}(\pi_{j|i})
     Q_{i|j}(\pi_{i|j})
         e^{-y\Delta F_{i j}}} \ .
    \end{eqnarray}

To characterize the statistical property of the vertex-cover problem
on a random graph, what we need is a distribution of the
distribution $Q_{j|i}(\pi_{j|i})$ among all the directed edges
$j\rightarrow i$ of the graph. Let us denote this distribution as
$\mathcal{P}_{1RSB}[Q(\pi)]$. Similar to Eq.~(\ref{eq:rs:pp}) we can
write down the following self-consistent equation for
$\mathcal{P}_{1RSB}[Q]$:
    \begin{equation}
        \label{eq:1rsb:pp}
        \mathcal{P}_{1RSB}[Q(\pi)]= p_{nn}(1)
         \delta\Bigl(Q(\pi)-\delta(\pi-\frac{e^{-\beta}}{e^{-\beta}+1})
         \Bigr) +
        \sum\limits_{k=1}^\infty p_{nn}(k+1)
        \int \prod_{j=1}^{k} \Bigl[ {\rm D} Q_j \mathcal{P}_{1RSB}(Q_j) \Bigr]
            \delta \Biggl(Q(\pi)-\mathcal{F}_{1RSB}(\{ Q_j\})  \Biggr) \ .
    \end{equation}
For graphs with  a general vertex degree distribution,
Eq.~(\ref{eq:1rsb:pp}) can be solved numerically by population
dynamics on a two-dimensional array (see, e.g., Ref.~\cite{Z08}). In
the special case of random regular graphs, the probability
distribution $\mathcal{P}_{1RSB}[Q]$ has a simple form:
$\mathcal{P}_{1RSB}[Q(\pi)]=\delta\bra{Q(\pi)-Q_c\bra{\pi}}$.  Then
Eq.~(\ref{eq:1rsb:pp}) can be re-written as a self-consistent
equation for a single  probability function $Q_c(\pi)$, and the
numerical task is much simplified.

\subsection{1RSB Stability analysis}
\label{subsec:1rsb-stability}

The stability of the 1RSB cavity solution is analyzed in the solution space
of the second-step replica-symmetry-breaking (2RSB) cavity theory.
In the 2RSB cavity theory, for each directed edge
$j\rightarrow i$ the order parameter
is the distribution of $Q_{j|i}(\pi_{j|i})$ over all the domains of
macrostates, which is denoted by
$\mathbb{Q}_{j|i}[Q]$. The iteration equation for this distribution reads
     \begin{equation}
     \label{eq:2rsb:iter}
        \mathbb{Q}_{j|i}[Q] = \frac{1}{\mathcal{Z}_{j|i}}
        \int \prod_{k\in \partial j \backslash i} D Q_{k|j}
     \mathbb{Q}_{k|j}[Q_{k|j}]
        e^{-y_2 \Delta G_{j|i} }
     \delta\Bigl(Q- \mathcal{F}_{1RSB}(\{Q_{k|j}\} )
     \Bigr) \,
    \end{equation}
where $\mathcal{Z}_{j|i}$ is a normalization constant, $y_2$ is the inverse
temperature at the level of
domains of macrostates, and $\Delta G_{k|j}$ is expressed as
    \begin{equation}
        \Delta G_{k|j} = -\frac{1}{y}\log\Bigl[\int\prod_{k\in\partial j
     \backslash i} {\rm d} \pi_{k|j}
            Q_{k|j}(\pi_{k|j}) e^{-y \Delta F_{k|j}} \Bigr] \ .
    \end{equation}
If on each directed edge $j\rightarrow i$ the iteration equation
Eq.~(\ref{eq:2rsb:iter}) converges to the
fixed-point solution
$\mathbb{Q}_{j|i}[Q(\pi)] = \delta\bigl(Q(\pi)-Q_{j|i}(\pi)\bigr)$, then the
1RSB solution is said to be stable toward further steps of
replica-symmetry-breaking.

According to Refs.~\cite{MR03,MPR04,RBMM04},
there are two types of instabilities the 1RSB cavity solution
Eq.~(\ref{eq:1rsb:iter}) can show toward non-trivial 2RSB
solutions. The first type of instability (type-I instability) is
state aggregation: the 1RSB macrostates aggregate into 2RSB domains,
while they themselves as described by Eq.~(\ref{eq:1rsb:iter})
contain no further internal structures.
The type-I instability can be studied
by tracing the propagation of a small perturbations to the
distribution $Q \bra{ \pi_{j|i} }$ during the 1RSB iteration. But in
practice it is rather difficult to implement such a check since the
distribution $Q_{j|i}(\pi)$ has to be represented by an array in the
numerical population dynamics simulation. In this paper, the type-I
instability analysis  is performed only at zero temperature for the
energetic cavity solution but not at finite temperatures.

The second type (type-II) instability is state fragmentation: a 1RSB
macrostate is itself composed of many sub-macrostates. Numerically,
this type of instability can be studied by tracing the propagation
of a small perturbation to $\pi_{i|j}$ during the 1RSB iteration.
The easiest way to do this is the deviation of two replicas method
\cite{Pagnani_etal_PRE_2003,ZDEB08}. One first iterates the
1RSB population dynamics to reach a steady state, and then creates a
replica of the whole population and gives a small perturbation to
each $\pi_{i|j}$ value of the origin population. These two
populations are then updated using the same sequence of random
numbers for a sufficiently long time. If the difference between the
two populations decays to zero with time, then the 1RSB cavity
solution is type-II stable. Another method of checking type-II
stability is noise propagation: we binds a noise $\chi_{j|i}$ to
each $\pi_{j|i}$ in the population. Then we iterate the population
using Eq.~(\ref{eq:1rsb:iter}) until a steady state is reached. At
the same time, the values of $\chi_{j|i}$'s are updated using
Eq.~(\ref{eq:stab:rs:iter}).  If $\sum_{j}\chi_{j|i}$ is decreasing
with iteration (equivalently, $\sum_{j}\chi_{j|i}<1$ finally), then
the 1RSB iteration is stable. We have checked both methods and find
that they always give the same result.

\subsubsection{The case of random regular graphs}

\begin{figure}
\includegraphics[width=0.7\textwidth]{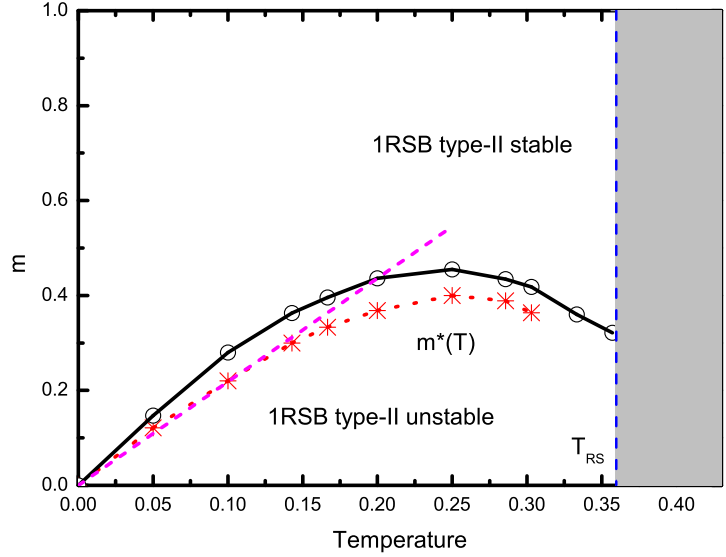}
\caption{ \label{fig:c5} (Color Online) Phase diagram of the 1RSB
solution of the vertex-cover problem on random regular graphs with
vertex degree $K=5$. When $T> T_{RS}(5)=T_d(5)=T_K(5) \approx 0.358$
the RS mean-field solution is stable (the shaded region) and 1RSB
solution at $m=1$ has only a trivial solution; when $T<T_{RS}(5)$, the
1RSB solution is type-II stable only when the Parisi parameter $m$
is located above the solid line which connects the circular symbols.
The values of the Parisi parameter $m=m^*$ which corresponds to zero
complexity and hence the dominating macroscopic states are given by
the dotted line. In this case, $m^*$ is always located in the
type-II unstable region. The dashed line represents the curve $m=y_I
T$, where $y_I$ is the maximal value of $y$ for which the 1RSB
zero-temperature energetic cavity solution is type-I stable.}
\end{figure}

\begin{figure}
\includegraphics[width=0.7\textwidth]{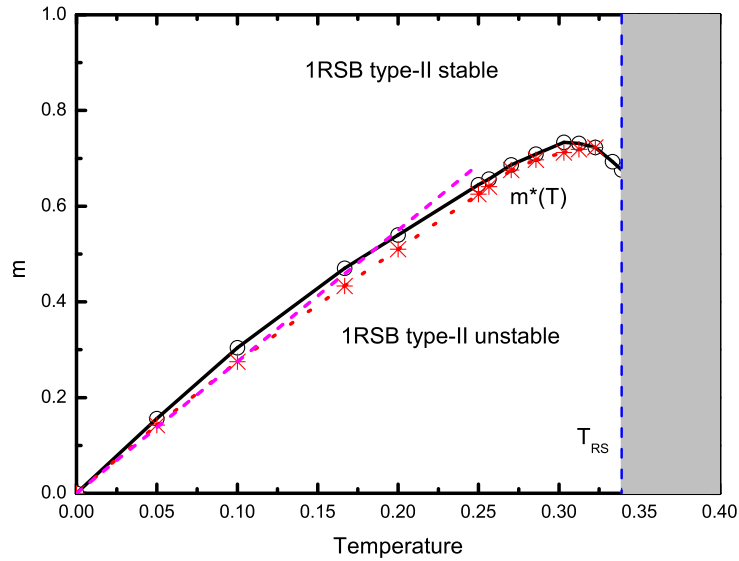}
\caption{ \label{fig:c10}
(Color Online) Same as Fig.~\ref{fig:c5}, but for random regular
graphs with vertex degree $K=10$.}
\end{figure}

Figure~\ref{fig:c5} shows the phase diagram for the random regular
graph VC problem with connectivity $K=5$. When temperature $T>
T_{RS}(K=5)=0.358$ the RS solution is stable. When $T< T_{RS}(5)$
the RS solution becomes unstable and $T_d=T_K=T_{RS}$. The 1RSB
solution is type-II stable only when the Parisi parameter $m$ is
sufficiently large. On the other hand, the physically meaningful
values of $m \leq m^*$, which correspond to $\Sigma(m) \geq 0$, are
all in the type-II unstable region (the value $m^*$ with
$\Sigma(m^*)=0$ as a function of $T$ is shown by the dotted line in
Fig.~\ref{fig:c5}). Therefore for  $K=5$ the 1RSB cavity solution is
insufficient to describe the statistical physics property of the VC
problem. The same qualitative results are obtained for random
regular graphs with $K=10$ (see Fig.~\ref{fig:c10}).

\begin{figure}
\includegraphics[width=0.7\textwidth]{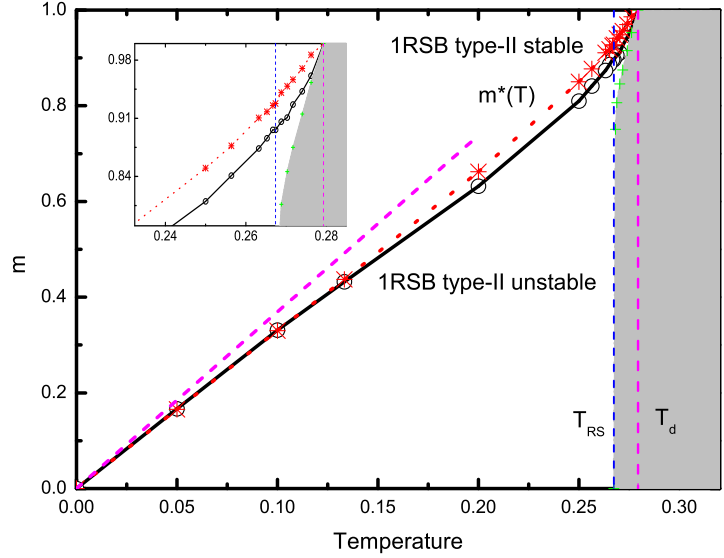}
\caption{
 \label{fig:c20}
 (Color Online) Same as Fig.~\ref{fig:c5},
but for random regular graphs with vertex degree $K=20$.  For
$T>T_{RS}(20)=0.2674$ the RS mean-field solution is locally stable, while
for $T<T_d(20) =T_K(20)= 0.2793$ a non-trivial 1RSB mean-field
solution appears at $m=1$. Notice that $T_d(20)> T_{RS}(20)$. The
boundary line (which connected the $+$ symbols) between the white
and the shaded regions marks the minimal value of $m$ below which
the 1RSB solution has no non-trivial solutions. The insert is an
enlarge of the main figure.}
\end{figure}

\begin{figure}
\includegraphics[width=0.7\textwidth]{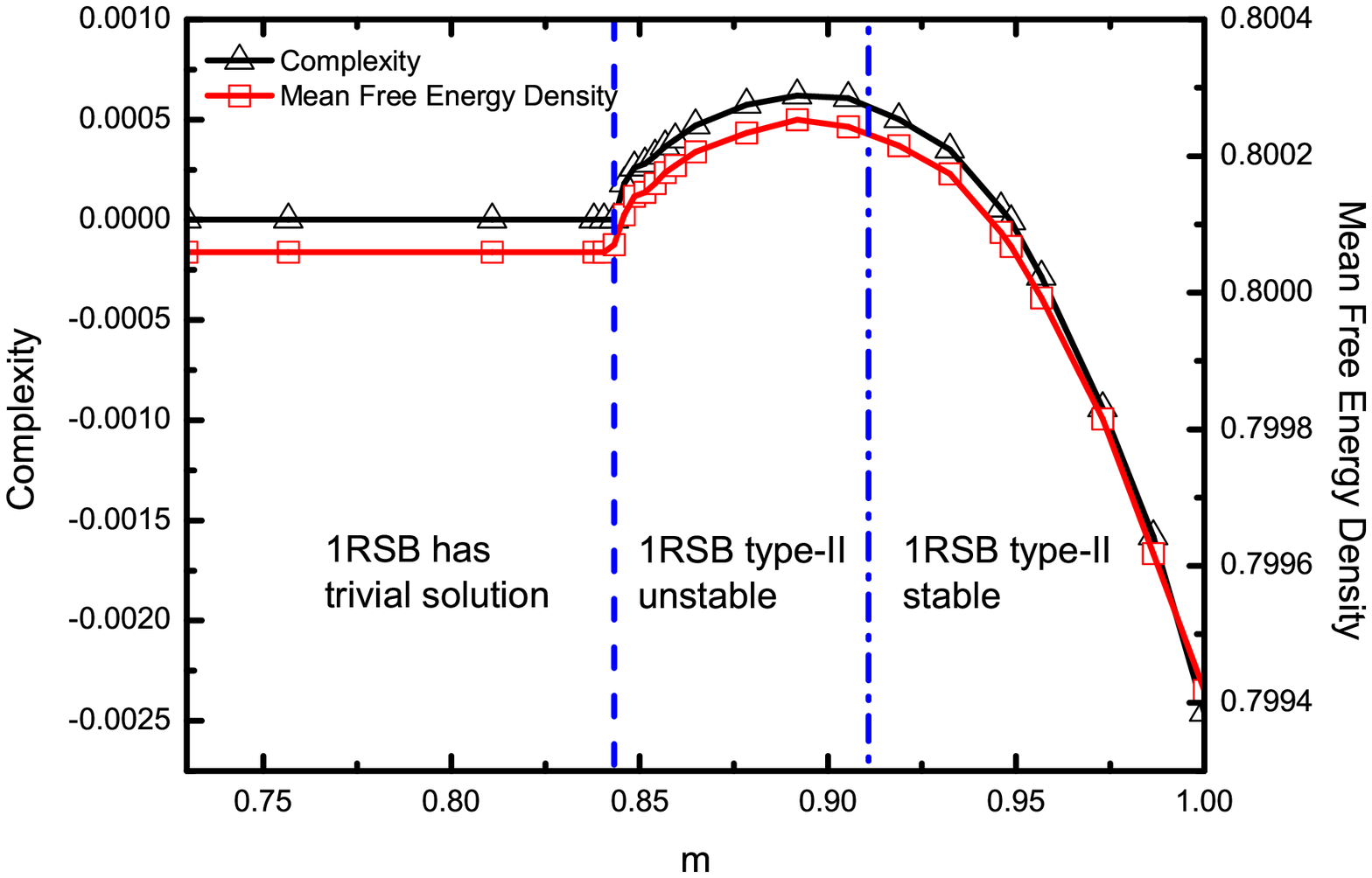}
\caption{ \label{fig:c20:com}
(Color Online) Complexity and mean free energy density at $T=0.2703$ for the VC problem on
regular graphs with vertex degree $K=20$.}
\end{figure}

Figure~\ref{fig:c20} shows the phase diagram for the random regular
graph vertex-cover problem with connectivity $K=20$. At this
connectivity we obtain results that are qualitatively different from
the results obtained for $K=5$ and $10$. The RS mean-field solution
is locally stable when temperature $T> T_{RS}(20)=0.2674$. In this case,
however, the dynamical transition temperature $T_d(K)$, which is
determined as the maximal temperature at which a nontrivial 1RSB
solution at $m=1$ exists, does not coincide with $T_{RS}(K)$.
 For $K=20$,
$T_{d}(20)=0.2793$, and in the temperature range $T_{RS}(20)< T <
T_d(20)$, non-trivial 1RSB solutions for the VC problem exist if the
Parisi parameter $m$ is beyond the boundary line between the white
and the gray region of Fig.~\ref{fig:c20}. This result indicates
that, in this temperature range, 1RSB solution and RS solution are
both stable. However the existence of a non-trivial 1RSB solution at
$m=1$ indicates that the RS solution is in fact not the physically
meaningful one, as it
does not describe the structure of the configuration space
correctly.

 As an example, for $K=20$ and $T=0.2703$, the
complexity and the free energy density of the 1RSB solution of the
VC problem are shown in Fig.~\ref{fig:c20:com} as a function of the
Parisi parameter $m$. When $m<0.8459$, the 1RSB solution reduces to
the RS solution, which has free energy density $f_{RS}\approx
0.80006$. A non-trivial 1RSB solution emerges for $m>0.8459$ and
this 1RSB solution becomes type-II stable when $m>0.9108$. The
complexity is a decreasing function of $m$ in the type-II stable
region and it reaches zero at $m=0.9486$ (correspondingly, the free
energy density of the dominating 1RSB macroscopic states is
$f_{1RSB}\approx 0.80008$). Therefore the 1RSB free-energy density
is only slightly larger than the RS free-energy density. For the
whole temperature range $T_{RS}(20) < T < T_d(20)$ we have checked
that the free energy density of the RS solution and that of the 1RSB
solution at $m=m^*(T)$ are always very close to each other.

Figure~\ref{fig:c20} also demonstrates that, when the temperature $T$
is higher than $0.15$, the line $m^*(T)$, which corresponds to the dominating
macroscopic states at each temperature, is located in the 1RSB type-II stable
region. If the 1RSB solution is also type-I stable at $m \approx m^*(T)$
(which we have checked to be the case only for $T=0$, see the dashed line),
then for $T > 0.15$ the VC problem can be sufficiently described by the 1RSB
solution without the need of further steps of replica-symmetry-breaking.
Further work is obviously needed to study more closely the
VC problem near the temperature $T_{RS}(K)$.
For very low temperatures, however, the 1RSB solution will become type-II
unstable.

\subsubsection{The case of random Erd\"{o}s-Renyi graphs}

\begin{figure}
\includegraphics[width=0.7\textwidth]{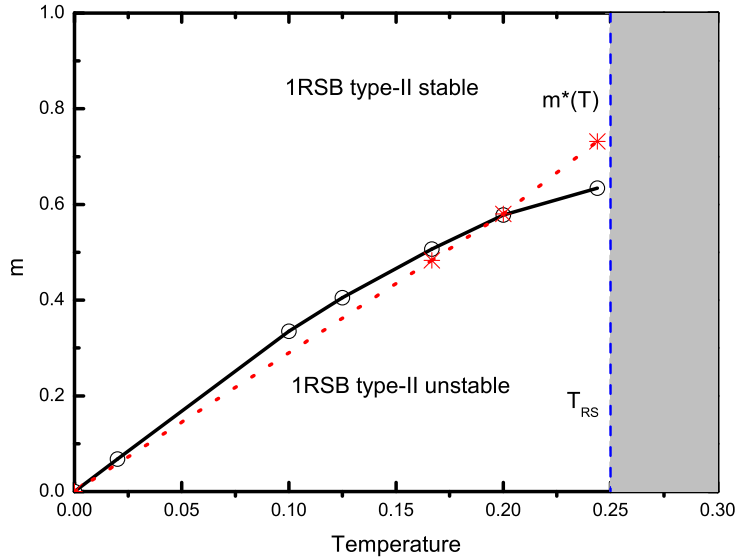}
\caption{ \label{fig:er:c10}
(Color Online)
 Phase diagram of the 1RSB solution of the VC problem on ER random graphs of
mean vertex-degree $c=10$. The RS solution is stable for temperature
$T>T_{RS}=T_{d}=T_K=0.25$.
For this system the energetic zero-temperature 1RSB solution is
type-I stable for $y \geq 0$.}
\end{figure}

Simulations on random ER graphs are technically more difficult, and
therefore we have studied only the cases of mean vertex degree $c=5$
and $c=10$. For the case of $c=5$, results similar to
Fig.~\ref{fig:c5} and Fig.~\ref{fig:c10} are obtained. The results
for the case of $c=10$ are shown in Fig.~\ref{fig:er:c10}. For this
system, the 1RSB solution at $m=m^*(T)$ is type-II stable when
$T>0.20$.

\subsection{The stability thresholds of the zero-temperature energetic
and entropic 1RSB cavity solution}
\label{subsec:zt:1rsb}

As a check of the finite-temperature results, here we compare the
low-temperature results with the results obtained directly at $T=0$.
At the zero temperature limit, two types of 1RSB solutions can be written down.
The energetic 1RSB solution \cite{Z03,WZ06}, which neglects all the entropic
effect of the VC problem,
is much simplified. Both the type-I and type-II stability analysis
of this solution can be performed. In this work, the  type-II stability
analysis is carried out through a bug proliferation
simulation (the detailed mathematical formulas are given in
Appendix~\ref{sec:appendixa}).

 The entropic 1RSB solution takes into account both the energetic and the
entropic effect and is numerically more involved.
For the vertex-cover problem, following the entropic zero-temperature RS
solution of Sec.~\ref{sec:rs:zerotemperature}, we can develop the 1RSB
solution by defining the 1RSB order parameter $Q_{j|i}(m_{j|i}, r_{j|i})$. The
iteration equation for $Q_{j|i}$ reads:
    \begin{equation}
     \label{eq:1rsb:zt:iter}
        Q_{j|i}(m, r) = \frac{1}{Z_{j|i}}
    \prod\limits_{k\in \partial j\backslash i} \Biggl[
        \sum_{m_{k|j}} \int {\rm d} r_{k|j} Q_{k|j}(m_{k|j}, r_{k|j}) \Biggr]
        e^{-y \Delta E_{j|i}} \delta \bigl(m-m_{j|i}\bigr)
        \delta\bigl(r-r_{j|i} \bigr) \ ,
    \end{equation}
where $m_{j|i}$ and $r_{j|i}$ are expressed by
Eq.~(\ref{eq:rs:eta-infty:m}) and Eq.~(\ref{eq:rs:eta-infty:r}).

When population dynamics is used to solve the entropic 1RSB equation
Eq.~(\ref{eq:1rsb:zt:iter}), it is observed that, if the re-weighting
parameter $y$ is lower than certain threshold value, the magnitudes of
some of the $r_{j|i}$ parameters may increase continuously with iteration
and eventually diverge. This divergence suggest that the zero-temperature
entropic 1RSB solution is not stable.
We use this divergence criterion to determine the type-II stability
threshold of the zero-temperature entropic 1RSB solution.

\begin{figure}
\includegraphics[width=0.9\textwidth]{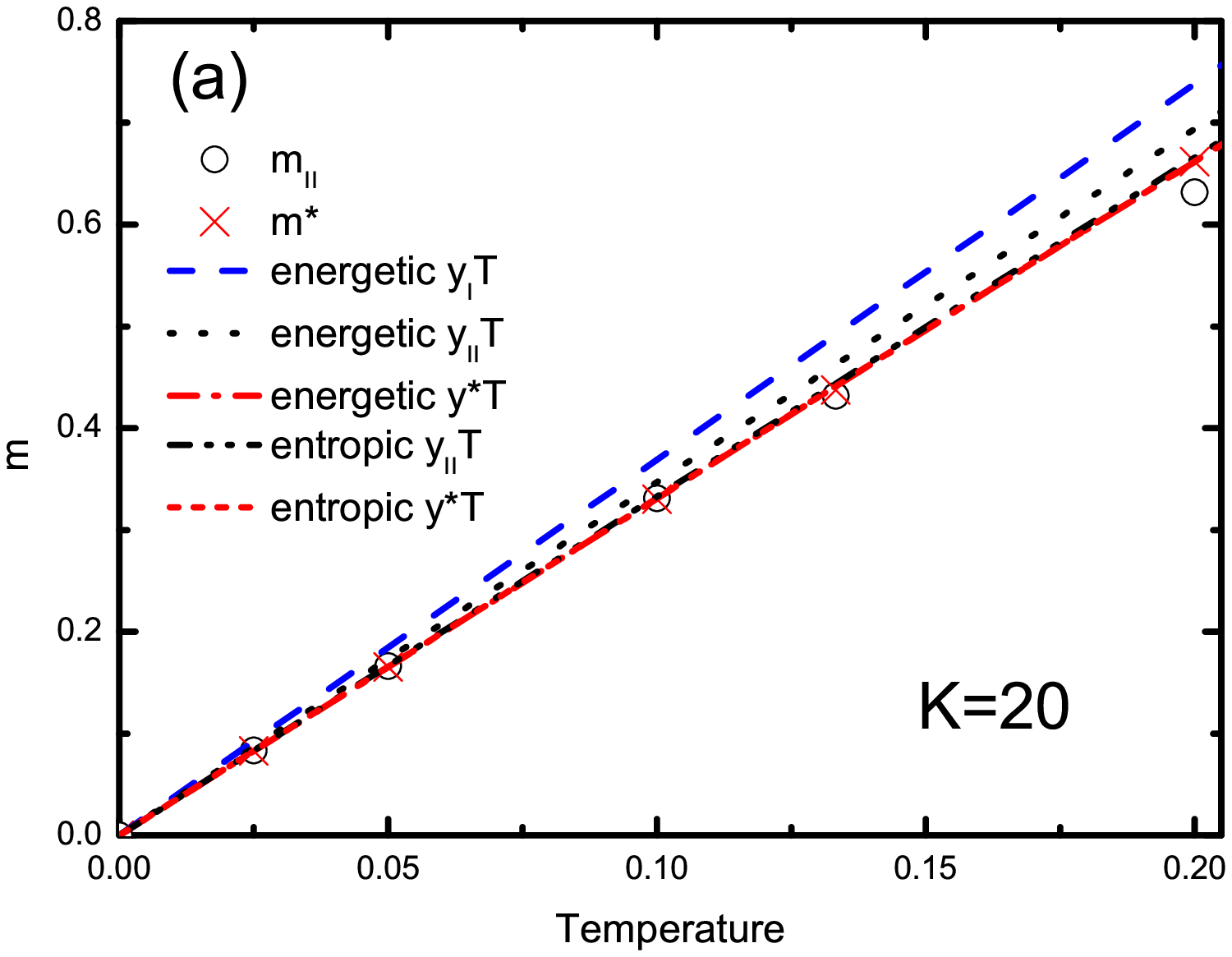}

\includegraphics[width=0.9\textwidth]{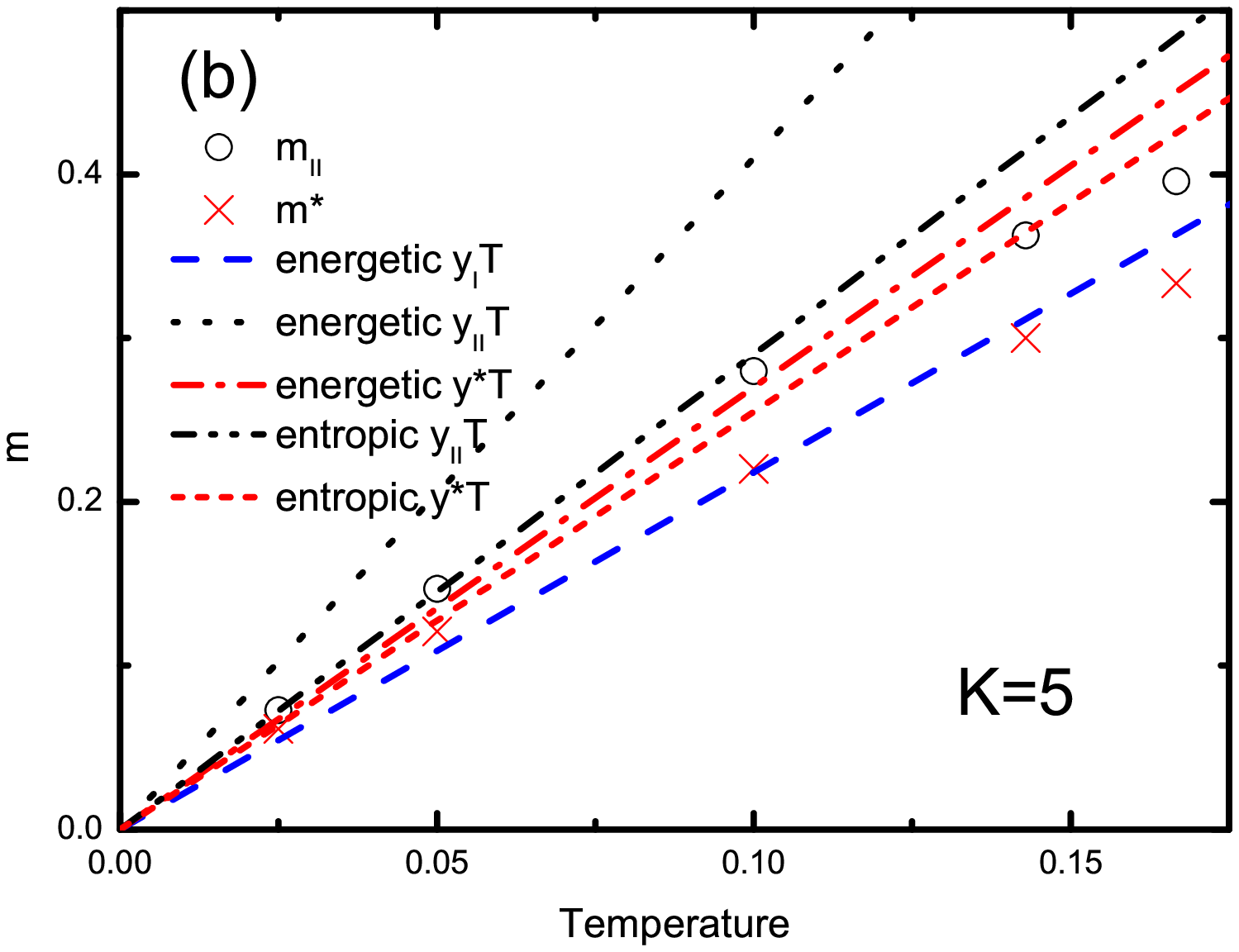}
\caption{
 \label{fig:zt:es}
(Color Online) Comparison of finite
temperature results with the zero-temperature energetic and entropic 1RSB
results for regular random graphs of vertex degree $K=20$ (a) and $K=5$ (b).
Symbols are finite-temperature results. The blue dashed curve ($y_I T$
represent the slopes corresponding of type-I stability of the energetic zero
temperature solutions. The energetic 1RSB solution is type-I stable when
$y<y_I$. Black curves (dotted and dash dotted) represent the slopes
corresponding to type-II stability of the (energetic and entropic)
zero-temperature 1RSB solution. The 1RSB solution is type-II stable when
$y>y_{II}$. Red curves (dash dotted and short dashed) represent the slopes
corresponding to $y(\Sigma=0)$ of the (energetic and entropic)
zero-temperature 1RSB solution.}
\end{figure}

Figure~\ref{fig:zt:es} shows the stability boundaries of the finite
temperature 1RSB solution, the energetic zero-temperature 1RSB
solution, and the zero-temperature entropic 1RSB solution,  for
random regular graphs with $K=20$ and $K=5$. For both $K=20$ and
$K=5$, the type-II stability threshold $y_{II}$ and value $y^*$
(determined by $\Sigma=0$) of the $T=0$ entropic 1RSB solution match
the corresponding slops in the $T$-$m$ plane of the
finite-temperature 1RSB solution.  At $K=20$ the $T=0$ energetic
1RSB solution has the same value of $y^*$ as that of the entropic
1RSB solution; and the type-II stability threshold $y_{II}$ of the
energetic 1RSB solution is very close to that of the entropic 1RSB
solution.

The energetic 1RSB solution is stable for $y< y_{I}$.
Since $y_I>y^*$ at $K=20$, the zero-temperature 1RSB solutions at
are type-I stable $y=y^*$. However, $y_I<y^*$ for $K=5$, therefore
the zero-temperature 1RSB solutions are type-I unstable at $y=y^*$.
At this value of vertex connectivity, the type-II stability
threshold $y_{II}$ as obtained for the 1RSB energetic solution and
the 1RSB entropic solution are different.

\section{Conclusion}
\label{sec:conclude}

In summary, the vertex-cover problem on finite-connectivity random
graphs were studied in this paper by finite-temperature cavity
method at both the replica-symmetric and first-step
replica-symmetry-breaking level, and the stability of these
mean-field solutions were analyzed. We found that the local stability
boundary $T_{RS}$ for the RS solution and the dynamical transition
temperature $T_d$ show a re-entrant behavior with the connectivity
both in the case of random regular graphs and Poisson random graphs:
the threshold
temperature $T_{RS}$ and $T_d$ first increases with connectivity and
then decreases with connectivity. The reason for this re-entrant
behavior (which is absent in the random $Q$-coloring problem
\cite{KZ07}) is not yet clear. For random regular graphs with a
relatively large connectivity (e.g., $K=20$), there exists a
temperature region in which both the RS solution and the 1RSB
solutions with Parisi parameter $m$ close to unity are stable. This
point deserves to be studied further on single graphs by the
belief-propagation iteration process \cite{YFW05,HW05} using
different initial conditions. At relatively large connectivity, the
VC problem at not too low temperatures may be sufficiently described
by the 1RSB cavity solution without the need of further steps of
replica-symmetry-breaking. But at temperature close to zero, more
complicated mean-field solutions are needed
\cite{WH01mvfrg,WZ06,ZMZ07}.

\section*{Acknowledgements}

We thank Jie Zhou for helpful discussions and thank
Florent Krzakala and Martin Weigt for their critical comments on an earlier version of the manuscript.
 PZ was supported by a graduate-student visiting fellowship
from the Chinese Academy of Sciences. The population dynamics simulations were performed at the High
Performance Computing Center of Lanzhou University and on  the
HPC cluster of ITP-CAS.
This work was partially supported by the National Science Foundation of China
(Grant number 10774150) and the 973-Program of China
(Grant number 2007CB935903).

\begin{appendix}

\section{Calculation of the dynamical transition temperature $T_d$}
\label{sec:appendixb}

The dynamical transition temperature $T_d$ is defined as the highest
temperature for the 1RSB cavity equation at Parisi
parameter $m=1$ to have a non-trivial solution. To
calculate $T_d$, one may solve the 1RSB equation Eq.~(\ref{eq:1rsb:iter})
with $m=1$ using population dynamics, but numerically this is quite
demanding, as population of populations is needed and different
macrostates should be properly reweighted.
It was first noticed in Ref.~\cite{MM06} that the 1RSB equation at
$m=1$ may be solved without using populations of populations and
reweighting of macrostates, and this possibility of simplification
was exploited in various later studies
\cite{ZK07,MRS08,ZDEB08,Zdeborova_Mezard_PRL_2008,Zdeborova_Mezard_JSM_12_12004_2008}.
In this appendix, we follow Ref.~\cite{MRS08} to solve
Eq.~(\ref{eq:1rsb:iter}) at $y=\beta$ (i.e., $m=1$).

At $m=1$ the mean cavity cover ratio $\overline{\pi}_{j|i}
\equiv \int {\rm d} \pi_{j|i} Q_{j|i}( \pi_{j|i}) \pi_{j|i}$ satisfies the
following iteration equation
\begin{equation}
    \label{eq:meanpiji}
  \overline{\pi}_{j|i} =
\frac{\emb}{\emb+\prod_{k\in\partial j\backslash
  i}\overline\pi_{k|j}} \ ,
\end{equation}
which has the same form as the RS iteration equation (\ref{eq:rs:iter}).
Therefore, distribution of $\overline{\pi}_{j|i}$ among all the edges of
the graph is given by $\mathcal{P}_{RS}$, see Eq.~(\ref{eq:rs:Pcv}).
Define $Q\bra{\pi | \overline{\pi} }$ as the conditional probability that
the cavity cover ratio is equal to $\pi$ when its mean value is
$\overline{\pi}$.
We have
\begin{eqnarray}
 & & Q\bra{\pi|\overline\pi}\mathcal{P}_{RS}\bra{\overline\pi} \nonumber \\
 & & \; \;  = \int {\mathcal{D}} Q \mathcal{P}_{1RSB}\brb{Q}
    Q\bra{\pi}\delta\bra{\overline\pi-\int d\pi Q\bra{\pi}\pi}\nonumber \\
& & \; \; =\sum\limits_{k=0}^\infty p_{nn}(k+1)
\int \prod\limits_{j=1}^{k}{\mathcal{D}} Q_j
    \mathcal{P}_{1RSB}(Q_j) \frac{\int \prod_{j=1}^k {\rm d} \pi_j
    Q_j(\pi_j) (e^{-\beta} + \prod_j \pi_j)}{e^{-\beta} + \prod_j \overline{\pi}_j} \delta\bigl(\pi-\mathcal{F}_{RS}(\{ \pi_j \}) \bigr)
    \delta\bigl(\overline{\pi}-\mathcal{F}_{RS}(\{ \overline{\pi}_j\})
    \bigr) \nonumber \\
& & \; \; = \sum\limits_{k=0}^\infty p_{nn}(k+1)
\int \prod\limits_{j=1}^{k} {\rm d} \overline{\pi}_j
{\mathcal{P}}_{RS}(\overline{\pi}_j)
\delta\bigl(\overline{\pi}-{\mathcal{F}}_{RS}(\{\overline{\pi}_j\}) \bigr)
\frac{\int \prod_{j} {\rm d} \pi_j {Q}_j(\pi_j | \overline{\pi}_j)
    (e^{-\beta}+\prod_j \pi_j)}{e^{-\beta}+\prod_j \overline{\pi}_j}
\delta\bigl(\pi-\mathcal{F}_{RS}(\{\pi_j\}) \bigr)  \ .
\label{eq:1rsbm1iter}
\end{eqnarray}
In deriving Eq.~(\ref{eq:1rsbm1iter}), we have used the
identity that
${Q}_{j}(\pi | \overline{\pi})
\equiv \int {\mathcal{D}} Q_j \mathcal{P}_{1RSB}(Q_j | \overline{\pi}) Q_j
(\pi)$,
with $\mathcal{P}_{1RSB}(Q_j | \overline{\pi})$ being the
conditional probability of $Q_j$ given that the mean value of the cavity
cover ratio is $\overline{\pi}$. $\mathcal{P}_{1RSB}(Q_j | \overline{\pi})$
is related to $\mathcal{P}_{1RSB}(Q_j)$ by
\begin{equation}
\mathcal{P}_{1RSB}(Q_j | \overline{\pi}) =
    \frac{\mathcal{P}_{1RSB}(Q_j) \delta\bigl(\overline{\pi}-
    \int {\rm d} \pi Q_j(\pi) \pi\bigr)}{
    \int \mathcal{D} Q_j \mathcal{P}_{1RSB}(Q_j) \delta\bigl(
    \overline{\pi}-\int {\rm d}\pi Q_j(\pi) \pi \bigr)}
    =   \frac{\mathcal{P}_{1RSB}(Q_j) \delta\bigl(\overline{\pi}-
    \int {\rm d} \pi Q_j(\pi) \pi\bigr)}{
    \mathcal{P}_{RS}(\overline{\pi})} \ .
\end{equation}

To get rid of the reweighting term $(e^{-\beta} + \prod\limits_j \pi_j)$
in Eq.~(\ref{eq:1rsbm1iter}), we define
$Q_{\sigma_j}(\pi_j | \overline{\pi}_j)$ as the conditional distribution
that the cavity cover ratio of vertex $j$ is equal to $\pi_j$ given that
the mean cavity cover ratio of vertex $j$ is $\overline{\pi}_j$ and
that vertex $j$ is in the spin state $\sigma_j$. We have
\begin{equation}
Q_{\sigma_j}(\pi_j | \overline{\pi}_j)
\equiv \frac{\psi_{\sigma_j} Q(\pi_j | \overline{\pi}_j) }
    {\overline{\psi}_{\sigma_j} } \ ,
\end{equation}
where
$\psi_{\sigma_j}=(1 - \pi_j)\delta_{\sigma_j, 1}+ \pi_j \delta_{\sigma_j,-1}$
is the probability distribution of $\sigma_j$, and
$\overline{\psi}_{\sigma_j}
=(1-\overline{\pi}_j)\delta_{\sigma_j,1}+\overline{\pi}_j
\delta_{\sigma_j,-1}$.
Then Eq.~(\ref{eq:1rsbm1iter}) can be rewritten as
\begin{eqnarray}
Q(\pi | \overline{\pi}) \mathcal{P}_{RS}(\overline{\pi})
& = &
    \sum\limits_{k=0}^\infty p_{nn}(k+1)
    \int \prod\limits_{j=1}^k {\rm d} \overline{\pi}_j
\mathcal{P}_{RS}(\overline{\pi}_j) \delta(\overline{\pi}-
\mathcal{F}_{RS}(\{\overline{\pi}_j\})  \nonumber \\
    & & \times \sum\limits_{\sigma_i} \sum\limits_{\{\sigma_j\}}
    \frac{\delta_{\sigma_i,-1} e^{-\beta} +
    \delta_{\sigma_i,1} \prod_j  \overline{\pi}_j
    \delta_{\sigma_j,-1} }
    {e^{-\beta} + \prod_j \overline{\pi}_j} \times
    \int \prod\limits_{j=1}^{k} {\rm d} \pi_j
    Q_{\sigma_j}(\pi_j | \overline{\pi}_j)
     \delta(\pi- \mathcal{F}_{RS}(\{\pi_j\})) \ .
\label{eq:noreweighting}
\end{eqnarray}

From the above equation and the identity that
\begin{equation}
Q\bra{\pi|\overline\pi}=\sum_\sigma\overline\psi_\sigma
Q_\sigma\bra {\pi|\bar\pi} \ ,
\end{equation}
we obtain an iterative equation for $Q_{\sigma}(\pi| \overline{\pi})$:
\begin{eqnarray}
  \overline\psi_\sigma
     Q_\sigma\bra{\pi|\overline\pi}\mathcal{P}_{RS}\bra{\overline\pi}
    &=& \sum_{k=0}^\infty {p}_{nn}(k+1)
 \int\prod_{j=1}^k {\rm d}\overline\pi_{j}
    \mathcal{P}_{RS}\bra{\overline\pi_{j}}\delta\brb{\overline\pi -
   \mathcal F_{RS}\bra{\bre{\overline\pi_j}}} \times \nonumber \\
    & & \times \sum_{\bre{\sigma_j}}
    \frac{e^{-\beta}\delta_{\sigma,-1} + \delta_{\sigma,1}
    \prod_j \overline{\pi}_j \delta_{\sigma_j,-1} }
{e^{-\beta}+ \prod_j \overline{\pi}_j}
    \times \int\prod_{j=0}^k {\rm d}\pi_j
  Q_\sigma\bra{\pi_j|\overline\pi_j} \delta\brb{\pi-\mathcal{F}_{RS}\bra{\bre{\pi_j}}} \ .
\label{eq:1rsbm1}
\end{eqnarray}

According to Ref.~\cite{MM06}, the 1RSB cavity equation
Eq.~(\ref{eq:1rsb:iter}) at $m=1$ has a non-trivial fixed point if
Eq.~(\ref{eq:1rsbm1}) has a non-trivial solution with the initial
conditions $Q_1 (\pi=0 | \overline{\pi})=1$ and $Q_{-1}(\pi=1 |
\overline{\pi}) = 1$, see also
Refs.~\cite{MRS08,ZDEB08,Zdeborova_Mezard_PRL_2008,Zdeborova_Mezard_JSM_12_12004_2008}.

\section{energetic zero-temperature stability analysis}
\label{sec:appendixa}

In this section, bug proliferation is used to analyze the type-II
instability of 1-RSB solutions in vertex-cover problems. We
introduce an message, the so-called warning $u_{j\rightarrow i}$
sent from vertex $j$ to a neighbor $i$. If the vertex $j$ is
uncovered, it sends a warning $u_{j\rightarrow i}=1$ to $i$,
otherwise $u_{j\rightarrow i}=0.$ To do survey propagation,
$\pi^{(0)}_{i\mid j}$ ($\pi^{(1)}_{i\mid j}$) is used to represent
the probability that vertex $i$ is always uncovered (covered) when
$j$ is removed in a cluster. Similarly, $\pi^{(*)}_{i\mid j}$ is the
probability that $i$ is unfrozen in the above situation.
\begin{align}
\pi^{(0)}_{i\mid l} &= c^{-1}_{i\mid l}\prod_{j\in N(i)\setminus
l}(1-\pi^{(0)}_{j\mid i}),\\
\pi^{(*)}_{i\mid l} &= c^{-1}_{i\mid l}e^{-y}\sum_{j\in
N(i)\setminus l}\pi^{(0)}_{j\mid i}\prod_{j^{'}\in N(i)\setminus
\{j,l\}}(1-\pi^{(0)}_{j^{'}\mid i}),\\
\pi^{(1)}_{i\mid l} &= c^{-1}_{i\mid l}e^{-y}[1-\prod_{j\in
N(i)\setminus l}(1-\pi^{(0)}_{j\mid i})-\sum_{j\in N(i)\setminus
l}\pi^{(0)}_{j\mid i}\prod_{j^{'}\in N(i)\setminus
\{j,l\}}(1-\pi^{(0)}_{j^{'}\mid i})],\\
c_{i\mid l} &=e^{-y}[1-(1-e^{y})\prod_{j\in N(i)\setminus
l}(1-\pi^{(0)}_{j\mid i}].
\end{align}

To analyze the type-II instability, a "bug" is introduced and
propagated on a graph. Here "bug" means supposed that along edge $1$
the warning is $\beta_1$, we turn it to another type like $\beta_0$
with a very small probability $p^1_{\beta_1\rightarrow \beta_0}$.
After one iteration, this will induce a new "bug" $\gamma
\rightarrow \delta$ on edge $l$ as an output and the probability of
this situation is:

\begin{equation}
p^l_{\gamma \rightarrow \delta}
=\frac1{Z}\sum_{\begin{array}{l}
(\beta_1, \cdots, \beta_n
) \rightarrow \gamma \\
(\beta_0, \cdots, \beta_n )\rightarrow
\delta
\end{array}}
(p^1_{\beta_1\rightarrow \beta_0}\cdots
p^n_{\beta_n})\exp{(-y\Delta E')}
\end{equation}
Thus we can define a matrix:
\begin{equation}
V_{\gamma\rightarrow\delta, \beta_1\rightarrow \beta_0}\equiv
\frac{\partial p^l_{\gamma \rightarrow \delta}}{\partial
p^1_{\beta_1\rightarrow \beta_0}}.
\end{equation}

The bug is propagated on the graph and if it can proliferate the
system is unstable. After $d$ times of iterations, absolutely the
criterion of such an instability is determined by a product of $d$
matrices
\begin{equation}
C\cdot |\lambda_{MAX}| \equiv \mu_d
\end{equation}
where $\lambda_{MAX}$ is the largest eigenvalue of matrix $V^1
\ldots V^d$. If $\mu_d$ grows exponentially with $d$, the solution
is unstable otherwise it is stable. Here the matrix $V$ is simply
just $2\times 2$,
\begin{equation}
\left( {\begin{array}{*{20}c}
   {V_{0 \to 1,0 \to 1} } & {V_{0 \to 1,1 \to 0} }  \\
   {V_{1 \to 0,0 \to 1} } & {V_{1 \to 0,1 \to 0} }  \\
\end{array}} \right)
\end{equation}

It is easy to get that:
\begin{align}
 V_{0 \to 1,1 \to 0} &=\frac{e^{y}\prod_{k\in N(i)\setminus {j,l}}(1-\pi^{(0)}_{k\mid j})}{1-(1-e^{y})\prod_{k\in N(i)\setminus j}(1-\pi^{(0)}_{k\mid
 j})}\\
V_{1 \to 0,0 \to 1} &=\frac{\prod_{k\in N(i)\setminus
{j,l}}(1-\pi^{(0)}_{k\mid j})}{1-(1-e^{y})\prod_{k\in N(i)\setminus
j}(1-\pi^{(0)}_{k\mid j})}
\end{align}
and $V_{0 \to 1,0 \to 1}=V_{1 \to 0,1 \to 0}=0$.

\begin{figure}
\label{fig:stab:zt}
\includegraphics[width=0.7\textwidth]{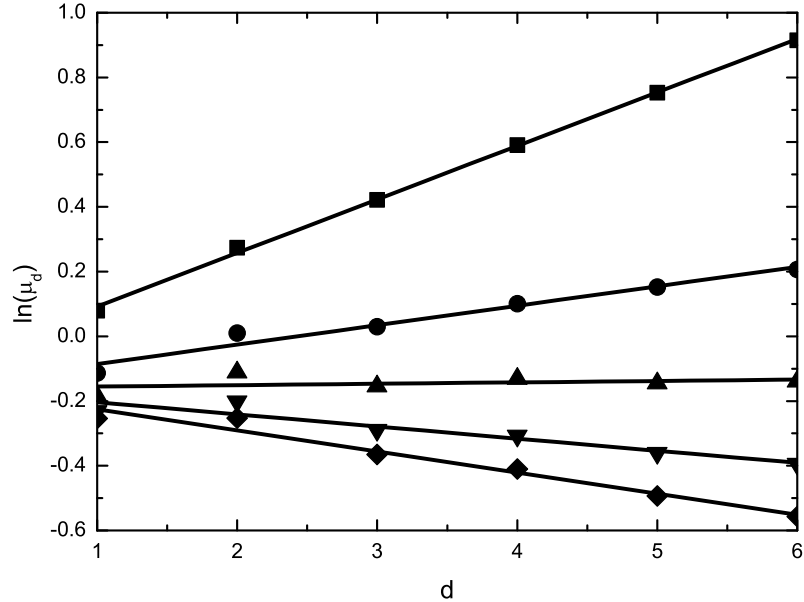}
\caption{Stability of the 1RSB solution of random vertex-cover
problem. $\ln \mu_d$ is plotted versus $d$ for different $y$. From
top to bottom: 2, 3, 3.3, 3.5, 3.55 and the lines are the linear
fits.}
\end{figure}

Solution about above equations and $\mu_d$ can be obtained through
population dynamics. We have applied this analysis on the 1RSB
solution of the vertex-cover problem with mean vertex degree $c=10$.
The results are shown in the fig.~\ref{fig:stab:zt}. We estimate the
threshold $y_{II}\approx 3.301$. Considering that the grand free-energy
density reaches the peak at $y^*=3.13$, we therefore conclude that
thermodynamics of the energetic zero-temperature 1RSB solution of the vertex-cover problem is unstable.

\end{appendix}

\bibliography{zp}

\end{document}